\documentclass[a4paper, amsfonts, amssymb, amsmath, reprint, showkeys, nofootinbib, twoside, superscriptaddress]{revtex4-1}\usepackage[english]{babel} 
\usepackage[utf8]{inputenc}
\usepackage[colorinlistoftodos, color=green!40, prependcaption]{todonotes}
\usepackage{ulem}
\usepackage{amsthm}
\usepackage{mathtools}
\usepackage{physics}
\usepackage{xcolor}
\usepackage{graphicx}
\usepackage[left=23mm,right=13mm,top=35mm,columnsep=15pt]{geometry} 
\usepackage{adjustbox}
\usepackage{placeins}
\usepackage[T1]{fontenc}
\usepackage{lipsum}
\usepackage{csquotes}
\usepackage[pdftex, pdftitle={Article}, pdfauthor={Author}]{hyperref}

\usepackage{xr}
\makeatletter
\newcommand*{\addFileDependency}[1]{
  \typeout{(#1)}
  \@addtofilelist{#1}
  \IfFileExists{#1}{}{\typeout{No file #1.}}
}
\makeatother

\newcommand*{\myexternaldocument}[1]{
    \externaldocument{#1}
    \addFileDependency{#1.tex}
    \addFileDependency{#1.aux}
}

\myexternaldocument{supp}

\begin{document}
%TC:ignore
\title{Probing Impurity Quantum Criticality with Entanglement Witnesses}
\newcommand{\mcw}[1]{{\color{blue}{#1}}}
\newcommand{\MCW}[1]{{\color{blue}{#1}}}
\newcommand{\ajm}[1]{{\color{red}{#1}}}
\newcommand{\AJM}[1]{{\color{red}{#1}}}
\renewcommand{\figurename}{FIG.}
\newenvironment{thisnote}{\par\color{purple}}{\par}

\author{Mateo Cárdenes Wuttig}
    \email[]{mateo.cardeneswuttig@yale.edu}
    \affiliation{Department of Applied Physics, Yale University, New Haven, Connecticut 06520, USA}

\author{Andrew J. Millis}
    \affiliation{Department of Physics, Columbia University, New York, New York 10027, USA}
    \affiliation{Center for Computational Quantum Physics, The Flatiron Institute, 162 5th Avenue, New York, New York 10010, USA}
\date{\today}

\begin{abstract}
Entanglement is a defining feature of quantum mechanics, and its relation to quantum criticality is of considerable current interest. Here we show that measurable spin and charge fluctuations provide an entanglement witness of quantum criticality in the two-impurity Kondo model, which has experimental realizations in terms of coupled quantum dots and magnetic impurities added to surfaces.
We use density-matrix renormalization group and numerical renormalization group calculations to resolve the non-Fermi-liquid critical point separating two independently Kondo-screened impurities from an inter-impurity singlet and interpret it as a change in the dominant entanglement partner of each local moment: from entanglement of the local moment with an extended set of conduction-electron degrees of freedom to entanglement with the other impurity. This reorganization is accompanied by a singular response of the impurity-bath entanglement and the inter-impurity susceptibility. We show how the same structure is encoded in the quantum Fisher information of collective spin and charge operators at zero and finite temperatures, connecting the entanglement picture to experimentally accessible dynamical response functions. Our results establish impurity systems as controlled settings in which quantum-critical entanglement can be detected through measurable correlations, and provide further insight into the possibility of understanding heavy-fermion physics in terms of entanglement.
\end{abstract}

\maketitle
%TC:endignore

% Content of paper
%%%%%%%%%%%%%%%%%%%%%%%%%%%%%%%%%%%%%%%%%%%%%%%%%%%%%%%%%
%\section{Introduction}\label{sec:outline}
%%%%%%%%%%%%%%%%%%%%%%%%%%%%%%%%%%%%%%%%%%%%%%%%%%%%%%%%%
\section{Introduction} The connection of quantum criticality to the reorganization of many-body entanglement is an important open problem. Here we study this issue in the context of the two-impurity Kondo problem\cite{PhysRevLett.47.737}, which consists of two spin-1/2 moments, each coupled by an exchange interaction $J$ to an independent bath of non-interacting conduction electrons and coupled to one another by an inter-impurity exchange $K$. The model has relevance to heavy fermion physics, magnetic impurities, and nanostructures. Closely related systems can be realized with coupled quantum-dot devices\cite{PhysRevLett.94.086602,PhysRevLett.108.086405,10.1038/s41567-022-01905-4,doi:10.1126/science.1095452}, in pairs of magnetic atoms assembled and probed by scanning tunneling microscopy (STM)\cite{10.1038/ncomms10046,nphys2076}, and with organic molecules on metallic surfaces\cite{PhysRevLett.103.087205,jacs.5c17416}.  
Related aspects of quantum-critical entanglement have previously been explored in the single-impurity, two-channel Kondo model \cite{sorensen2007,PhysRevB.93.081106}, which has the same universal critical behavior \cite{PhysRevLett.108.086405} as the two-impurity model studied here. 
Our work also connects to recent work on the relationship between spin-conduction electron entanglement and multipartite entanglement at the Kondo-destruction critical point\cite{ncomms2498,fang2026quantumfisherinformationmagnetic}.

For antiferromagnetic couplings ($J,K>0$) and in the absence of charge transfer between the two baths, the model has a non-Fermi-liquid quantum critical line $K_c(J)$ separating a phase where the impurities are screened by conduction electrons forming Kondo screening clouds\cite{PhysRevB.80.205114,PhysRevLett.86.2854} from a phase where the two impurities are locked in a singlet state\cite{PhysRevLett.61.125,PhysRevB.80.205114,PhysRevB.52.9528,PhysRevB.40.324,PhysRevLett.68.1046}.
The two phases differ by the dominant entanglement partner of each local moment: for $K<K_c(J)$, each impurity forms an extended many-body singlet with its own conduction bath, whereas for $K>K_c(J)$, the impurities form an almost pure mutual singlet and become weakly entangled with the baths\cite{PhysRevLett.109.066403,ncomms4784}. We show in this controlled impurity quantum-critical setting that experimentally accessible spin and charge fluctuations can provide insight into the changes of entanglement across quantum phase transitions. 

Entanglement entropies are not directly accessible through conventional measurements, which motivates the use of entanglement witnesses\cite{https://doi.org/10.1002/qute.202400196}. In this paper, we study the quantum Fisher information (QFI)\cite{PhysRevA.85.022322}, which can be expressed in terms of experimentally accessible local fluctuations and cross-correlations, or equivalently reconstructed from the corresponding frequency-resolved susceptibilities. At zero temperature, the QFI is completely determined by the quantum fluctuations of the collective operators, while at finite temperature, the dissipative dynamical response function isolates quantum from thermal fluctuations\cite{nphys3700}.  
This provides a direct route to connect the system's response to local magnetic fields or potential differences to entanglement and many-body correlations. 

Here, we use the density-matrix renormalization group (DMRG) algorithm in matrix product state formalism\cite{PhysRevLett.69.2863,dmrg_uli} to determine the zero-temperature phase diagram and obtain representations of the wave functions, from which entanglement and the corresponding witnesses can be computed. By combining the impurity correlations, entanglement entropies, and the critical response, we resolve the non-Fermi-liquid critical point separating the Kondo-screened and inter-impurity-singlet phases and extend the analysis into the ferromagnetic sectors of the phase diagram. We show that the QFI of local impurity-spin and adjacent-electron density operators tracks the redistribution of entanglement across these transitions\cite{PhysRevLett.109.066403, ncomms4784} and allows certification and characterization of entanglement across the whole parameter space. A finite-temperature treatment is performed with the numerical renormalization group (NRG) algorithm\cite{RevModPhys.47.773,PhysRevB.79.085106} to determine the experimental window over which the non-Fermi-liquid crossover remains visible in the dynamical response and to characterize the energy scale up to which the QFI can certify entanglement. 

%
%The remainder of this paper is organized as follows. Section~II introduces the two-impurity Kondo model, its schematic phase diagram, and the entanglement and QFI diagnostics. Section~III presents our results: Sec.~III\,A maps correlations, entanglement, and QFI across the zero-temperature phase diagram; Sec.~III\,B shows how collective spin and charge fluctuations reveal the non-Fermi-liquid critical point and the ferromagnetic-sector transitions; and Sec.~III\,C extends the analysis to finite temperatures and determines the range over which signatures of criticality and entanglement persist in dynamical response functions. Section~IV summarizes our conclusions and discusses the experimental implications.

%%%%%%%%%%%%%%%%%%%%%%%%%%%%%%%%%%%%%%%%%%%%%%%%%%%%%%%%%
\section{Model and entanglement witnesses}\label{sec:methods}
%%%%%%%%%%%%%%%%%%%%%%%%%%%%%%%%%%%%%%%%%%%%%%%%%%%%%%%%%
\begin{figure*}[htb]
 \centering
 \begin{adjustbox}{center}
 %0.75
   \includegraphics[width=0.75\textwidth]{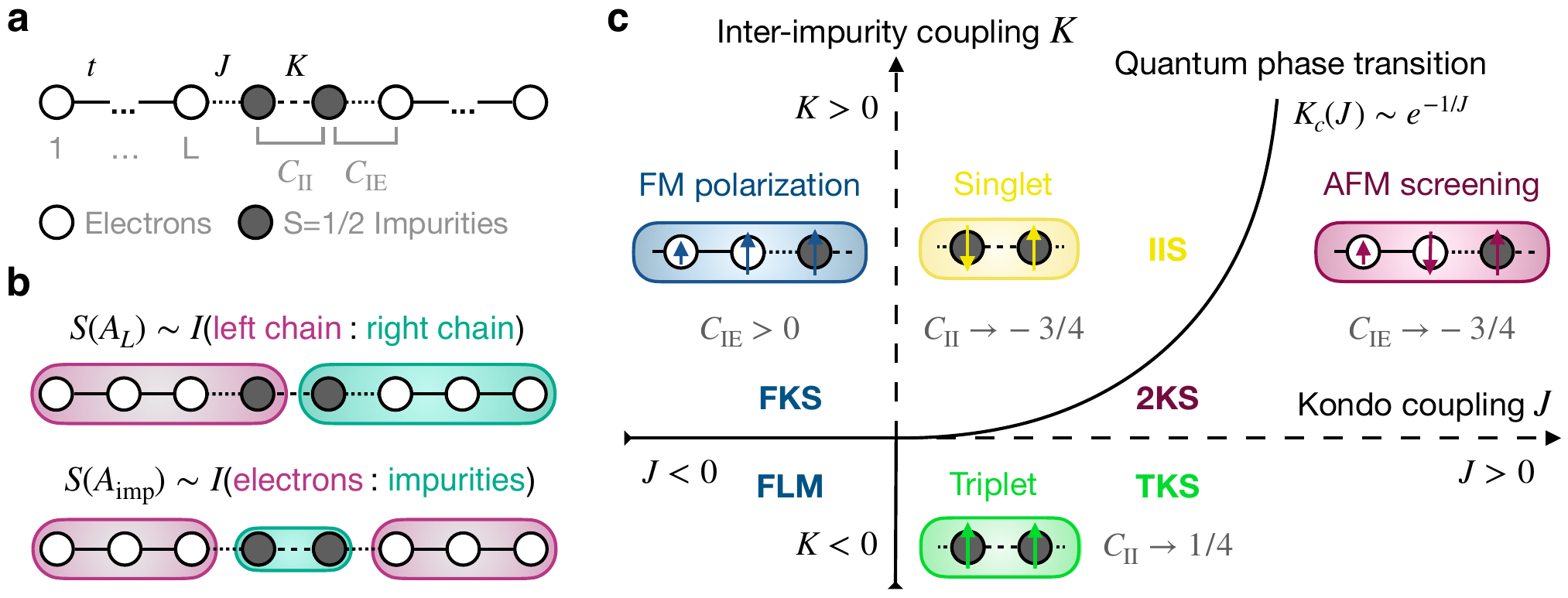} %1.5
 \end{adjustbox}
 \caption{\textbf{Two-impurity Kondo model, entanglement partitions, and schematic phase diagram.}
\textbf{a}, Two spin-$1/2$ impurities (gray sites) are coupled by an inter-impurity exchange $K$ and connected through Kondo couplings $J$ to independent half-filled chains of $L$ non-interacting electrons (white sites) with nearest-neighbor hopping $t$. $C_{\rm II}=\langle\mathbf{S}_1\cdot\mathbf{S}_2\rangle$ and $C_{\rm IE}=\langle\mathbf{S}\cdot\mathbf{s}_{{\rm e}}\rangle$ denote the impurity-impurity and impurity-electron correlators, respectively.
\textbf{b}, Partitions used to probe ground-state entanglement, which is proportional to the mutual information $I$: Two single-impurity chains $S(A_L)$ (top) and impurities and electrons $S(A_{\rm imp})$ (bottom).
\textbf{c}, Schematic zero-temperature ground-state phase diagram as a function of $J$ (horizontal axis) and $K$ (vertical axis).
Along $K=0$, the system reduces to two independent single-impurity Kondo models with antiferromagnetic (AFM) screening and ferromagnetic (FM) polarization for $J>0$ and $J<0$, respectively.
For $J=0$, the impurities decouple from the electron baths and form a singlet (triplet) for $K>0$ ($K<0$).
Finite AFM couplings produce the inter-impurity singlet (IIS) and two-Kondo-singlet (2KS) phases, separated by a critical line $K_c(J) \sim e^{-1/J}$.
The FM sector consists of triplet Kondo-screened (TKS), FM local-moment (FLM), and FM-polarized singlet-dominated (FKS) regimes.
Solid and dashed lines denote quantum phase transitions and crossovers, respectively.}
 \label{fig:system_and_phase_diagram}
\end{figure*}

Our model consists of two spin $S=1/2$ impurities coupled via a Heisenberg exchange. Each of the impurities is also connected to an independent electron chain of length $L$, as sketched in Fig. \ref{fig:system_and_phase_diagram}(a). The Hamiltonian is
\begin{equation}
    H = H_{\rm e} + H_{\rm imp} + H_{\rm Kondo} \,,
\end{equation}
where $H_{\rm e} = -t\sum_{\alpha} \sum_{\langle i,j \rangle, \sigma} (c^\dag_{i,\alpha,\sigma} c_{j,\alpha,\sigma} + \text{h.c.})$ describes two half-filled, non-interacting electron chains $\alpha = 1,2$ with hopping amplitude $t$ between nearest-neighbor sites $\langle i,j \rangle$. The operator $c^\dag_{i,\alpha,\sigma}$ ($c_{i,\alpha,\sigma}$) creates (destroys) an electron with spin $\sigma = \uparrow, \downarrow$ on site $i$. 
The inter-impurity coupling $K$ between the two impurity spins is 
\begin{equation}
    H_{\rm imp} = K \, \mathbf{S}_1 \cdot \mathbf{S}_2 \,,
\end{equation}
and the Kondo exchange coupling $J$ is
\begin{equation}
    H_{\rm Kondo} = J \,\sum_{\alpha} \mathbf{S}_{\alpha} \cdot \mathbf{s}_{{\rm e},\alpha} \,.
\end{equation}
Here, $\mathbf{S}_{\alpha}$ consists of conventional Heisenberg spin operators, and $\mathbf{s}_{{\rm e},\alpha}$ denotes the electron spin on the endpoint of the chain adjacent to impurity $\alpha$.
We use the von Neumann entropies $S(A_L)$ and $S(A_{\mathrm{imp}})$ for the partitions in Fig. \ref{fig:system_and_phase_diagram}(b), measuring left-right and impurity-bath entanglement, respectively\cite{dmrg_uli}. The former includes correlations crossing the central cut, while the latter measures impurity-bath entanglement.

\subsection{Phase diagram}\label{sec:phase_diagram}
Fig.~\ref{fig:system_and_phase_diagram}(c) shows the schematic phase diagram of this model, summarizing the competition between the inter--impurity exchange $K$ and the Kondo coupling $J$ to the conduction electrons. The central diagnostics are the impurity--impurity correlator $C_{\rm II}= \langle \mathbf{S}_1 \cdot \mathbf{S}_2 \rangle$
and the impurity--electron correlator $C^{\alpha}_{\rm IE} = \langle \mathbf{S}_{\alpha} \cdot \mathbf{s}_{\rm e,\alpha} \rangle$.

At $K=0$ (horizontal axis), antiferromagnetic (AFM) Kondo coupling, $J>0$, drives each impurity towards a Kondo-screened Fermi-liquid state, characterized by $C_{\rm IE}<0$, consistent with local AFM screening correlations $C_{\rm IE}\to -3/4$ in the strong local-coupling limit $J/t \rightarrow \infty$. AFM screening leads to a short screening length in the strong-coupling limit and an increasing screening-cloud size for decreasing $J>0$\cite{PhysRevLett.109.066403}. For $J<0$, the ferromagnetic (FM) Kondo coupling leads to an unscreened local moment with positive local impurity--electron correlations $C_{\rm IE}>0$.
% first sector
At nonzero $J$ and $K$, the phase diagram contains distinct phases and crossover regimes. 
In the two-Kondo-singlet phase (2KS), realized for $J>0$ and $K\lesssim T_K(J)$, each impurity is screened by its own electron bath, such that $C_{\rm IE}<0$ and $C_{\rm II}\rightarrow 0$. 
For $K \gtrsim T_K(J)$, the impurities instead bind into an inter-impurity singlet (IIS), such that $C_{\rm II}\to -3/4$ and $C_{\rm IE}\to 0$.
The boundary between the 2KS and IIS phases is a quantum critical curve\cite{PhysRevB.40.324} (solid line in Fig. \ref{fig:system_and_phase_diagram}c) controlled by the competition $K = \mathcal{O}[ T_K(J)]$. The Kondo temperature of this model is estimated as\cite{PWAnderson_1970,DONIACH1977231}
$
    T_K(J) \sim 2t \exp\left(-\frac{\pi t}{2J}\right)\,,
$
with a convention-dependent prefactor. For fixed $J>0$, we define $K_c(J)$ as the critical inter-impurity coupling at which the phase transition occurs. 
The corresponding single-impurity Kondo length\cite{PhysRevB.80.205114,PhysRevLett.86.2854} is $\xi_K\sim T_K^{-1}$. Finite temperature $T$ and finite chain length provide temporal and spatial infrared cutoffs, respectively. The Kondo-screened state and its extended impurity-bath entanglement are only fully resolved when $T \ll T_K$ or equivalently $L\gg\xi_K$. Because $T_K$ is exponentially small at weak $J$, the temperature required to observe Kondo screening decreases, and the necessary system size increases, exponentially as $J\rightarrow0^+$. At finite temperature, the zero-temperature quantum phase transitions are rounded into crossover regimes. Near the critical point in the sector $J,K>0$, the low-energy flow is governed by the non-Fermi-liquid fixed point over the window $T^*\ll T\ll T_K$, where $T^*\propto (K-K_c)^2/T_K$. 

% entanglement in first sectors
At $K=0$, $S(A_L)$ is zero; as $K$ increases, the exchange coupling generates entanglement between the two subsystems, and $S(A_{\rm L}) > 0$. In the large-$K$ limit, the impurities form an almost pure singlet, which is cut by the bipartition, and $S(A_{\rm L})\rightarrow \ln 2$. The opposite trend is expected for $S(A_{\rm imp})$. In the Kondo-screened limit at small $K$, each impurity is entangled with its adjacent electron chain; the impurity-electron entanglement approaches $S(A_{\rm imp})\rightarrow2\ln2$, the maximum permitted by the four-dimensional impurity Hilbert space.
As $K$ increases, the impurities increasingly bind to each other and the impurity-electron entanglement decreases. For $K \gg J$, the impurity subsystem becomes approximately a pure singlet, and $S(A_{\rm imp})\rightarrow 0$. Thus, $\partial_K S(A_{\rm imp})$ vanishes as $K \rightarrow 0,\infty$, and is expected to be maximal near the critical coupling $K_c$. 

 % other sectors 
For $J>0$ and sufficiently negative $K<0$, the impurities form a triplet with $S(A_{\rm imp})\rightarrow \ln{3}$, which is screened by the two conduction channels, giving a triplet Kondo-screened regime (TKS) with $C_{\rm II}\to 1/4$ and $C_{\rm IE}<0$. This regime is adiabatically connected to the 2KS regime across $K=0$ and separated from it by a crossover (dashed line in Fig. \ref{fig:system_and_phase_diagram}c). For $J<0$ and $K<0$, both couplings favor ferromagnetic alignment, producing a FM local-moment regime (FLM) with $C_{\rm II}\to 1/4$ and FM polarization, $C_{\rm IE}>0$. Finally, for $J<0$ and $K>0$, the impurity singlet competes with local FM impurity--electron alignment. For weak $|J|$, the state remains singlet dominated, $C_{\rm II}\approx -3/4$, while increasing $|J|$ produces a FM polarized, singlet-dominated regime (FKS), characterized by a reduced magnitude of $C_{\rm II}$ and $C_{\rm IE}>0$.
The two ferromagnetic-sector boundaries are quantum phase transitions corresponding to a Fermi-liquid singlet-triplet level crossing at $K=0$ for fixed $J<0$ and an AFM-FM Kondo boundary for an effective spin-one triplet at $J=0$ for fixed $K<0$.

\subsection{Quantum Fisher information}\label{sec:entanglement_witnesses}
For a thermal state $\rho_T$ at temperature $T$, the quantum Fisher information (QFI) of a generator $\mathcal{O}$ is obtained from the imaginary, dissipative part of the dynamical susceptibility\cite{nphys3700}
\begin{equation}\label{eq:qfi_response}
    F_Q[\rho_T,\mathcal{O}]
    =
    \frac{4}{\pi}
    \int_0^\infty \dd{\omega}\,
    \tanh\left(\frac{\omega}{2T}\right)
    \chi''_{\mathcal{O}\mathcal{O}}(\omega,T),
\end{equation}
where $\chi''_{\mathcal{O}\mathcal{O}}(\omega,T)$ is the corresponding response function. If the system is in a pure state $\ket{\psi}$, the QFI is proportional to the variance, 
\begin{equation}
    F_Q[\ket{\psi},\mathcal{O}] = 4\,\text{Var}(\mathcal{O}) \,,
\end{equation}
while $F_Q[\rho_T,\mathcal{O}] \le 4 \text{Var}_{\rho_T}(\mathcal{O})$ for a mixed state.
For $\mathcal{O} = \mathcal{O}_L \pm \mathcal{O}_R$ and two local operators such as $S^z$ or $n_{\sigma}$ which are normalized to unit spectral width, every state separable across $L \vert R$ obeys $F_Q \leq 2$, and the maximum value is $F_Q=4$. Hence, $F_Q > 2$ witnesses left-right entanglement\cite{PhysRevA.85.022321}.
The Methods sections gives the operator definitions and reconstruction from local and cross responses. DMRG yields the zero-temperature phase diagram and collective-operator variances, and NRG the finite-temperature dynamical response; implementation and convergence details are in the supplementary material.

%%%%%%%%%%%%%%%%%%%%%%%%%%%%%%%%%%%%%%%%%%%%%%%%%%%%%%%%%
\section{Entanglement across the phase diagram}\label{sec:results_phase_diagram}
%%%%%%%%%%%%%%%%%%%%%%%%%%%%%%%%%%%%%%%%%%%%%%%%%%%%%%%%%
% Figure 2: Main result -- phase diagram
\begin{figure*}[htb!]
 \centering
 \begin{adjustbox}{center}
   \includegraphics[width=0.7\textwidth]{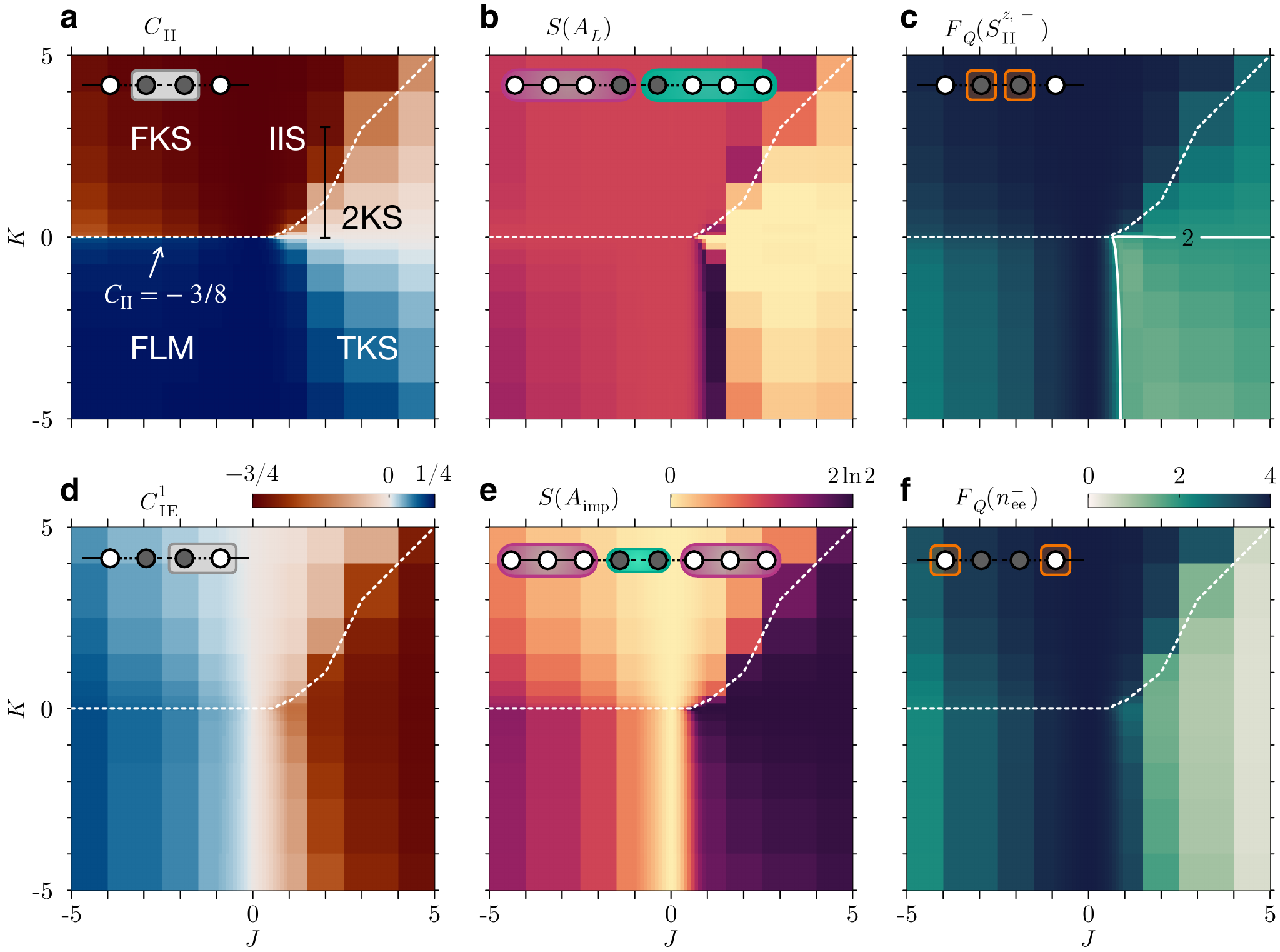} %1.4
 \end{adjustbox}
\caption{\textbf{Correlations, entanglement, and quantum Fisher information across the phase diagram.}
Zero temperature ground-state observables as a function of the electron--impurity Kondo coupling $J$ (horizontal axis) and inter-impurity exchange $K$ (vertical axis), obtained from DMRG.
\textbf{a}, \textbf{d}, Spin-spin correlations.
The impurity--impurity correlator
$C_{\rm II}$
(\textbf{a}) distinguishes predominantly singlet-correlated (red) and triplet-correlated (blue) impurities.
In all panels, the dotted white contour marks $C_{\rm II}=-3/8$ and provides a reference scale for the competition between singlet formation and Kondo screening in the quadrant $J,K>0$. The black line corresponds to the linecut presented in Fig. 3.
The impurity--electron correlator
$C^{1}_{\rm IE}$
(\textbf{d}) measures the correlation between the left impurity and the nearest electron site in its bath; positive (blue) and negative (red) values indicate local ferromagnetic polarization and antiferromagnetic screening, respectively.
\textbf{b}, \textbf{e}, Entanglement entropy for the partitions illustrated in the insets: (\textbf{b}) between the two single-impurity subsystems $S(A_L)$, which approaches $\ln2$ in the isolated inter-impurity-singlet limit and is reduced when the impurities become uncorrelated, and (\textbf{e}) between the impurities and conduction electrons $S(A_{\rm imp})$, which is largest in the Kondo-screened regimes and vanishes when the impurity-electron coupling becomes zero.
\textbf{c}, \textbf{f}, Quantum Fisher information
$F_Q(S^{z,-}_{\rm II})$ and $F_Q(n^{-}_{\rm ee})$
for the collective impurity spin-spin operator $S^{z,-}_{\rm II} = S^z_1 - S^z _2$ and density imbalance 
$n^{-}_{{\rm ee}} = n_{{\rm e},1} - n_{{\rm e},2}$ of the closest conduction electrons, respectively. The solid white contour in panel \textbf{c} marks $F_Q=2$, which is the separability bound involving the impurity-spin degrees of freedom. Both QFIs track the inter-impurity and impurity-electron correlation, but only a spin-resolved QFI (\textbf{c}) also verifies entanglement.
}
 \label{fig:phase_diagram}
\end{figure*}

Figure \ref{fig:phase_diagram} summarizes the zero-temperature DMRG phase diagram in units of $t=1$. 
In the antiferromagnetic sector, $J,K>0$, the competition between Kondo screening and inter-impurity singlet formation is revealed by the impurity-impurity correlator $C_{\rm II}$, see Fig.~\ref{fig:phase_diagram}a. The impurity-electron correlator $C_{\rm IE}$ in Fig.~\ref{fig:phase_diagram}d provides a complementary diagnostic to $C_{\rm II}$, distinguishing Kondo screening $C_{\rm IE} < 0$ (red) for $J>0$ from ferromagnetic polarization $C_{\rm IE} > 0$ (blue) for $J<0$ throughout the phase diagram; for $\abs{J} \rightarrow 0$, the correlations are reduced ($C_{\rm IE} \rightarrow 0$).
For large $J>0$, each impurity is screened predominantly by its own bath, where $C_{\rm II} \rightarrow 0$ and $C_{\rm IE} \rightarrow -3/4$ in the two-Kondo-singlet (2KS) phase, whereas increasing $K$ binds the impurities into an inter-impurity singlet (IIS) with $C_{\rm II} \rightarrow -3/4$ and $C_{\rm IE} \rightarrow 0$. These phases are separated by the established non-Fermi-liquid critical point\cite{PhysRevLett.61.125,PhysRevB.40.324,PhysRevB.52.9528}. 
The remaining quadrants correspond to ferromagnetic regimes. For $K<0$, the impurities develop triplet correlations $C_{\rm II} > 0$ (blue), which coexist with AFM impurity--electron correlations $C_{\rm IE} < 0$ for $J>0$, yielding a triplet Kondo-screened regime (TKS), corresponding to an effective two-bath, single spin-1-impurity problem, and with ferromagnetic bath polarization $C_{\rm IE} > 0$ for $J<0$, characteristic of a local-moment regime (FLM). For $J>0$, the inter-impurity correlations vanish for $\abs{K} \rightarrow 0$. For $J<0$ and $K>0$, the inter-impurity singlet formation competes with the weak ferromagnetic polarization of the surrounding electrons (FKS). 

Next, we discuss the redistribution of entanglement between impurities and baths. 
Fig.~\ref{fig:phase_diagram}b shows the entropy $S(A_L$) between the two single-impurity subsystems, which is enhanced $S(A_L)\to\ln2$ when the impurities are correlated across the central bond and suppressed when each impurity is predominantly screened within its own bath. The bright region corresponds to large Kondo screening $C_{\rm IE} \rightarrow -3/4$ with vanishing inter-impurity correlations $C_{\rm II} \rightarrow 0$ and $S(A_L) \to 0$. By contrast, the impurity-electron entropy $S(A_{\rm imp})$ shown in Fig.~\ref{fig:phase_diagram}e is largest in the Kondo-screened regimes: it approaches  $S(A_{\mathrm{imp}})\to2\ln2$ in the 2KS limit and $\ln 3$ in the triplet-screened limit, and is reduced to zero when the impurity subsystem becomes factorized from the baths, $C_{\rm IE} \rightarrow 0$ for $\abs{J} \rightarrow 0$. 

The same structure is encoded in the QFI of experimentally accessible collective-spin and charge fluctuations. Fig.~\ref{fig:phase_diagram}c presents the
antisymmetric channel $F_Q(S^{z,-}_{\rm II})$, which is enhanced in singlet-correlated regions. The contour $F_Q=2$ (white line) indicates where the corresponding two-particle separability bound is exceeded, indicating left–right entanglement between the two impurity–bath subsystems. 
The bath-density QFI shows that the rearrangement of quantum correlations extends beyond the impurity sector and is detectable through local fluctuations of the surrounding conduction electrons (Fig.~\ref{fig:phase_diagram}f); it is largest for $J \rightarrow 0$, and decreases to zero when the impurities are AFM screened in the TKS and 2KS phases. The computed $F_Q(n^{-}_{\rm ee})$ remains below the left-right separability bound $F_Q = 8$, consistent with local charge QFI in related bulk Fermi liquids\cite{wang2025localnonlocalentanglementwitnesses}, but still tracks impurity–electron correlations.

A spin-resolved bath-density QFI yields a phase diagram similar to that of $F_Q(n^{\pm}_{\rm ee})$, where the symmetric and antisymmetric components of $F_Q(n^{\pm}_{\sigma,\rm ee})$ provide complementary certification of entanglement across the phase diagram. 
Thus, the QFI serves two complementary roles: it certifies entanglement when a separability bound is violated, which requires a spin-resolved measurement, and it probes spin-spin correlations and entanglement redistribution when tracked as a function of the coupling parameters, even when it does not witness entanglement.
% Reference to supplementary material
The supplementary material shows that the same QFI also resolves the quantum phase transitions across the ferromagnetic-sector boundaries, witnesses entanglement, and tracks its redistribution.

\section{Local response at the quantum critical point} %\subsection
% Figure 3: critical point and QFI
\begin{figure*}[htb!]
 \centering
 \begin{adjustbox}{center}
   \includegraphics[width=1\textwidth]{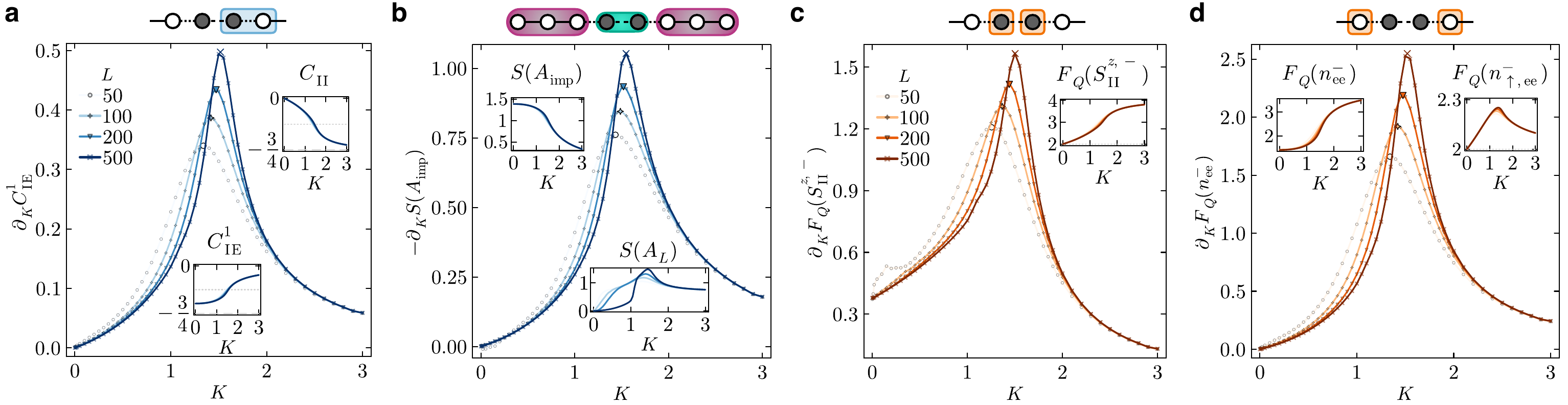} %2
 \end{adjustbox}
\caption{\textbf{Quantum Fisher information reveals entanglement redistribution at the critical point.}
Derivatives with respect to the inter-impurity exchange $K$ for $J=2$ and different electron chain lengths $L$.
\textbf{a}, The derivative of the impurity--electron correlator,
$\partial_K C^1_{\rm IE}$, develops a peak near the size-dependent critical coupling $K_c(L)$, visualized by large symbols. The insets show $C^1_{\rm IE}$ and the impurity--impurity correlator $C_{\rm II}$.
\textbf{b}, The derivative of the entropy between electrons and impurities, $-\partial_K S(A_{\rm imp})$, provides an entanglement signature of the critical point. The insets show $S(A_{\rm imp})$ and $S(A_L)$, demonstrating how entanglement is redistributed across the phase transition.
\textbf{c}, The derivative of the QFI, $\partial_K F_Q(S^{z,-}_{\rm II})$, is proportional to $\partial_K C_{\rm II}$ and recasts criticality into an experimentally measurable quantity. 
\textbf{d}, The derivative
$\partial_K F_Q(n^-_{{\rm ee}})$
of the electron density imbalance on the closest conduction electron sites provides an independent probe of the critical point.
The insets in \textbf{c} and \textbf{d} show the corresponding undifferentiated QFI, including a spin-resolved QFI $F_Q(n^-_{{\uparrow,\rm ee}})$ (inset in \textbf{d}).
$F_Q(S^{z,-}_{\rm II}),F_Q(n^-_{{\uparrow,\rm ee}}) > 2$ verify entanglement for all $K>0$.
Panels \textbf{a} and \textbf{b} present the known impurity quantum critical point using conventional correlation and entanglement measures, while panels \textbf{c} and \textbf{d} demonstrate that it can also be detected through the QFI of collective impurity-spin and local electron-density operators.
}
 \label{fig:critical_point}
\end{figure*}

Fig.~\ref{fig:critical_point} shows how the critical response at the non-Fermi-liquid critical point is encoded in correlation functions, entanglement, and the QFI. At fixed impurity-electron coupling $J=2$, the derivatives of $C_{\rm II}$ and $C^1_{\rm IE}$ with respect to $K$ develop pronounced maxima\cite{PhysRevLett.109.066403, RevModPhys.80.517, https://doi.org/10.1038/416608a} at a system-size-dependent coupling $K_c(L)$, see Fig.~\ref{fig:critical_point}a. 
Resolving the critical point requires  $L \gg \xi_K$, with $\xi_K \sim e^{1/J}$ growing exponentially with decreasing $J$.
The negative derivative of $S(A_{\rm imp})$ peaks in the same region (Fig.~\ref{fig:critical_point}b), and $S(A_L)$ is maximal (inset), reflecting the redistribution of quantum correlations as the ground state evolves from predominantly Kondo-screened impurities for small $K < K_c$ to an inter-impurity singlet with increasing inter-impurity coupling $K > K_c$. 
The staggered impurity susceptibility $\chi'({S^{z,-}_{\rm II}})$ peaks at the critical point, as shown in Fig. \ref{fig:NRG_QFI}a, and the derivatives $\partial_K C_{\rm IE}$, $\partial_K S(A_{\rm imp})$, and $\partial_K F_Q(S^{z,-}_{\rm II})$ become non-analytic, even though the respective correlators themselves remain smooth because both adjacent phases are singlets\cite{PhysRevB.96.041109}. 

The behavior of the susceptibilities is very similar to that found in studies of the Bose-Fermi impurity model\cite{cai2019dynamicalkondoeffectkondo}; however, in a lattice model, the entanglement at the ordering wavevector was found to peak at the critical point\cite{ncomms2498}.
In the supplementary material Sec. \ref{sec:kondoscreeninglength}, we present the correlation-weight screening length $\xi^1$, which develops a size-dependent maximum near the pseudo-critical coupling, showing that the screening cloud extends increasingly far into the baths before being cut off by finite size. The peak positions of $\partial_K C_{\rm II}$ extrapolate to $K_c = 1.620 \pm 0.005$, consistent with the $T=0.001$ NRG value $1.59 \pm 0.03$.

The same critical response is directly visible in $F_Q(S_{\rm II}^{z,-})$, see Fig.~\ref{fig:critical_point}c.
Note that for the SU(2)-invariant state with vanishing impurity magnetization, $F_Q(S^{z,\pm}_{\rm II}) = 2 \pm \frac{8}{3}C_{\rm II}$, recasting the impurity correlations in an interpretable entanglement witness\cite{ncomms2498}. 
For $K > K_c$, $C_{\rm II} \to -3/4$, so the antisymmetric QFI saturates its two-impurity maximum $F_Q(S^{z,-}_{\rm II}) \to 4$, while the symmetric channel vanishes.
In contrast, $F_Q(n^{-}_{{\rm ee}})$ provides an independent probe of the critical response. Fig. \ref{fig:critical_point}d shows that $\partial_K F_Q(n^-_{{\rm ee}})$ develops a pronounced peak near the same finite-size critical coupling $K_c(L)$. 
Its critical enhancement reveals that the redistribution of entanglement is imprinted on spatially resolved occupation fluctuations of the conduction electrons, providing an experimentally motivated route to detect impurity quantum criticality without direct measurement of the inter-impurity spin correlation. 
The spin-resolved QFI $F_Q(n^{-}_{\uparrow,{\rm ee}})$ (inset of Fig. \ref{fig:critical_point}d) follows the critical response and also verifies entanglement\cite{ncomms2498}. 

%%%%%%%%%%%%%%%%%%%%%%%%%%%%%%%%%%%%%%%%%%%%%%%%%%%%
\section{Thermal survival of entanglement signatures} %%%%%%%%%%%%%%%%%%%%%%%%%%%%%%%%%%%%%%%%%%%%%%%%%%%%

\begin{figure*}[htb!]
 \centering
 \begin{adjustbox}{center}
 %1a
   \includegraphics[width=1\textwidth]{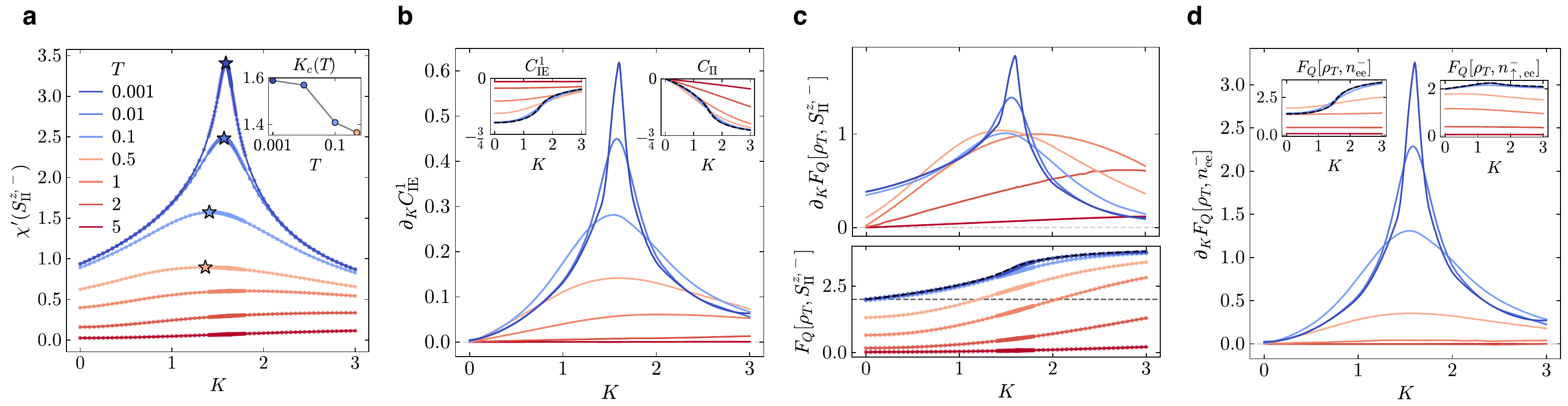} %2
 \end{adjustbox}
\caption{\textbf{Impurity criticality and QFI at finite temperatures.}
\textbf{a}, Static susceptibility $\chi'(S^{z,-}_{\rm II})$ of the staggered impurity spin operator as a function of $K$ for $J=2$ for different temperatures $T$. All data are obtained from NRG. The low-temperature peak at $T=0.001$ identifies $K_c = 1.59 \pm 0.03$ and is suppressed into a broader crossover with increasing temperature. 
\textbf{b}, Derivative of the impurity--electron correlator $\partial_K C^1_{\rm IE}$, whose extrema track the susceptibility peak. Insets show the corresponding finite-temperature correlators, where $T=0$ DMRG data (dashed black lines) of the undifferentiated observables show that both methods agree in the limit $T \to 0$. 
\textbf{c,d}, QFI of the impurity generator $S^{z,-}_{\rm II}$ and of the conduction-electron generator $n^{-}_{\rm ee}$. The derivatives are enhanced near the critical region and decrease with increasing temperature. The undifferentiated impurity-spin QFI in \textbf{c} demonstrates that entanglement can be certified up to $T \approx 0.5$.}
\label{fig:NRG_QFI}
\end{figure*}

At finite temperature, the thermal von Neumann entropy contains quantum and classical contributions and is therefore not an entanglement measure. The dissipative QFI response (eq. \ref{eq:qfi_response}) removes the classical contribution and witnesses the remaining quantum correlations. We present NRG data for several temperatures $T$.
Fig. \ref{fig:NRG_QFI}(a) shows the static susceptibility $\chi'({S^{z,-}_{\rm II}})$ of the critical operator as a function of $K$ for $J=2$, which peaks around $K_c = 1.59 \pm 0.03$ at $T=0.001$. This uncertainty reflects the finite-temperature scale and grid spacing.
The critical response of the closest conduction electrons $\chi'({s^{z,-}_{\rm ee}})$ peaks at the same critical coupling $K_{c, {\rm ee}} = 1.59 \pm 0.03$ (supplementary material Sec. \ref{sec:NRG_dynamical_quantities}), explaining the non-analytic behavior of observables on the closest conduction electrons close to the critical point. 
Increasing temperature broadens and suppresses these features as the flow leaves the critical regime, which survives for $T^*\ll T\ll T_K$.
In fig. \ref{fig:NRG_QFI}(b), $\partial_K C^1_{\rm IE}$ tracks the susceptibility peak, and the correlators $C^1_{\mathrm{IE}}$ and $C_{\mathrm{II}}$ (insets) show agreement between the $T=0$ data (dashed black line) from DMRG and the $T=0.001$ NRG calculations.
At $K = 0$, the impurities are uncorrelated and $C_{\rm II} = 0$; increasing $K>0$ leads to singlet correlations $C_{\rm II} < 0$. At fixed $K$, increasing temperature leads to decreasing spin-spin correlations; however, for $0 < T < T^*$, we find that this trend is non-monotonic.
Similarly, at $K=0$, the correlator $C_{\rm IE} < 0$ indicates AFM screening of the single-impurity Kondo problems for $T \rightarrow 0$. Increasing temperature leads to free impurities with $C_{\rm IE} \rightarrow 0$ for $T \gg T_K$. 

At all temperatures, spin and electron-density QFIs follow the impurity correlations and critical response.
For $T \lesssim 0.01$, a sharp response of the finite-temperature crossover is visible; while for larger temperatures, $T \gtrsim 2$, the absence of structure in the QFI as a function of $K$ is consistent with thermally reduced correlations. 
The derivative $\partial_K F_Q[\rho_T,S^{z,-}_{\rm II}]$ allows identification of the critical point, see Fig. \ref{fig:NRG_QFI}(c). This QFI witnesses entanglement over a range of coupling strengths $K$ even at $T=1$; the separability bound (dotted line) is violated up to $T \sim 0.5$ at $K_c$. 
We note that temperatures on the order of the coupling terms are large and provide a promising route towards detecting the critical point and certifying entanglement in experiments. 
The adjacent-electron density QFI shows that the impurity fluctuations imprinted on the conduction electrons survive the same temperature range in which the crossover is visible, see Fig. \ref{fig:NRG_QFI}(d). The spin-resolved QFI (inset) witnesses entanglement up to $T \sim 0.1$, which is lower than the impurity QFI in panel (c). 
NRG shows that a common low-energy scale controls the divergent impurity susceptibility, finite-temperature peak collapse, and loss of QFI certification.
This makes the finite-temperature QFI a practical guide for experiments, fixing both the response channel to measure and the temperature range in which a fluctuation spectrum can still be interpreted as a witness of impurity quantum-critical entanglement.
The supplementary material presents a minimal, exactly solvable four-spin strong-coupling limit that reproduces local entanglement and its collective-spin QFI certification up to temperatures set by the exchange coupling, and explains the finite-temperature behavior of $C_{\rm II}$ and $C_{\rm IE}$ in Fig. \ref{fig:NRG_QFI}(b).

%%%%%%%%%%%%%%%%%%%%%%%%%%%%%%%%%%%%%%%%%%%%%%%%%%%%%%%%%
\section{Experimental implementation and outlook}\label{sec:discussion}
%%%%%%%%%%%%%%%%%%%%%%%%%%%%%%%%%%%%%%%%%%%%%%%%%%%%%%%%%
Coupled quantum dots provide realizations of two-impurity Kondo systems \cite{PhysRevLett.94.086602,PhysRevLett.108.086405,10.1038/s41567-022-01905-4,doi:10.1126/science.1095452} that naturally support measurements of the entanglement witnesses discussed here. Common-mode and differential gate-voltage modulations couple directly to charge operators of the form $n^{+}_{\rm ee}$ and $n^{-}_{\rm ee}$, respectively, while radio-frequency reflectometry measures the in-phase and quadrature components of the complex admittance and hence separates the dispersive and dissipative parts of the response\cite{PhysRevLett.108.036802}.
Surface-assembled magnetic atoms offer a complementary platform with atomic spatial resolution\cite{PhysRevLett.98.056601}. 
Scanning tunnelling spectroscopy (STM) can resolve local Kondo resonances and spin excitations of individual magnetic atoms and coupled atomic structures\cite{PhysRevLett.103.107203}. Attaching one atom to an STM tip allows one to tune continuously the inter-impurity coupling through the tip–sample separation\cite{nphys2076}.
Electron-spin-resonance STM further enables local coherent driving and readout of individual and exchange-coupled surface spins\cite{PhysRevLett.122.227203,doi:10.1126/science.abg8223,ncomm2023}. 
Together, these protocols enable detection of the impurity quantum critical point through experimentally accessible uniform and staggered responses, while bath-density observables track critical restructuring as it propagates into surrounding conduction electrons.

Several extensions are of interest. 
Charge transfer between the two baths rounds the critical point considered here into a crossover\cite{PhysRevB.91.155140,PhysRevLett.97.166802,PhysRevLett.108.086405}. Other realizations of the critical point are also unstable against particle-hole symmetry breaking\cite{PhysRevLett.61.125,PhysRevB.40.324,PhysRevB.52.9528}. Future work should quantify the evolution of entanglement and test the robustness of witnesses against unequal impurity couplings, particle-hole asymmetry, dissipation, and inter-bath charge transfer.
Spatially resolved electronic responses may further track how the critical reorganization propagates from the impurities into the Kondo screening cloud.
The two-impurity critical point shares the low-energy fixed point of the two-channel Kondo model and its residual thermodynamic impurity entropy of $\frac{1}{2}\ln 2$\cite{PhysRevLett.108.086405}, which is distinct from the impurity-electron entanglement entropy $S(A_{\rm imp})$ studied here.
In real-space studies of the one- and two-channel Kondo models, a bulk-subtracted block entropy resolves the screening cloud, decreasing from $\ln 2$ to $0$ for one-channel screening but only to $\frac{1}{2}\ln 2$ at the overscreened two-channel fixed point\cite{sorensen2007,PhysRevB.93.081106}. 
Whether spatially integrated staggered-spin QFI detects this crossover and screening-cloud entanglement is also an important question for further investigation.

Our two-impurity results offer a complementary microscopic benchmark for recent Kondo-lattice calculations and inelastic-neutron-scattering measurements, which have found strongly enhanced spin QFI near Kondo-destruction quantum criticality\cite{ncomms2498,Mazza2026}.
Staggered total-spin generators combining local moments with conduction-electron degrees of freedom, together with momentum-resolved spin and charge QFI, may connect this impurity-scale rearrangement to the formation and loss of coherent heavy quasiparticles and help distinguish Kondo-destruction dynamics from conventional order-parameter fluctuations.
Thus, the two-impurity Kondo problem provides not only a paradigmatic model of non-Fermi-liquid criticality, but also a benchmark for detecting quantum-critical entanglement through measurable fluctuations.

%%%%%%%%%%%%%%%%%%%%%%%%%%%%%%%%%%%%%%%%%%%%%%%%%%%%%%%%%
%TC:ignore
\section{Methods}
\subsection{Numerical simulations}\label{sec:dmrg}
Large-scale numerical simulations are performed over a wide range of physical parameters with the density matrix renormalization group (DMRG) algorithm\cite{PhysRevLett.69.2863} in matrix product state (MPS) formalism\cite{dmrg_uli} using the ITensor library\cite{itensor} to obtain zero-temperature ground state wave functions from which the correlation functions, entanglement entropies, and entanglement witnesses are computed. 
This method is particularly well suited for one-dimensional Hamiltonians with short-range interactions. 
The data presented in Fig. 2 are obtained with a bond dimension cutoff of $\chi_{\max} = 4000$ in a system with $L=500$ to reach convergence. Fig. 3 contains data with $\chi_{\max} = 4000$ for systems with different lengths between $L=50$ and $L=500$.
Due to a known even-odd finite size effect, we only use even lengths $L$ for each chain\cite{PhysRevB.53.9153, PhysRevLett.86.2854, PhysRevB.91.155140}. Additionally, we use the numerical renormalization group (NRG) 
algorithm in the NRG Ljubljana/TRIQS implementation\cite{PhysRevB.79.085106} to calculate finite-temperature spectral properties and correlations. 
Instead of a fixed length $L$, we use up to $N_{\rm keep} = 1500$ many-body multiplets. Details of the finite-size analysis and convergence checks are reported in the supplementary material.

\subsection{Entanglement witnesses}
We characterize entanglement via the von Neumann entropy between two connected subsystems $A$ and $B$, defined as
$
        S(\rho_A) = - \Tr(\rho_A \log{\rho_A}) = S(\rho_B)
$\cite{dmrg_uli},
where $\rho_A$ and $\rho_B$ denote the reduced density matrices of the ground state in the respective region. In a pure state, the mutual information is $I(A:B) = 2 S(A)$. The partitions in  Fig. \ref{fig:system_and_phase_diagram}(b) are used to calculate the entanglement entropy $S(A_{\rm L})$ between the two single-impurity chains, where $A_{\rm L}=\{\text{left electrons}+I_1\}$ is the region of the left chain with impurity $I_1$, and $S(A_{\rm imp})$ between electrons and impurities $A_{\rm imp}=\{I_1,I_2\}$. $S(A_{\mathrm{imp}})$ measures impurity-bath entanglement, but does not by itself determine how this entanglement is distributed spatially within the baths; $S(A_L)$ includes the contribution of an inter-impurity singlet crossing the central cut, together with any additional many-body entanglement generated between the two conduction channels.

To witness the entanglement, we consider the collective impurity-spin generators
$
    S^{z,\pm}_{\rm II} = S^z_{\rm 1} \pm S^z_{2} 
$.
Here, $S^{z,+}_{\rm II}$ probes uniform, ferromagnetic fluctuations, while $S^{z,-}_{\rm II}$ detects staggered, antiferromagnetic fluctuations. The corresponding QFI can be probed with an oscillating magnetic field applied with the same or opposite sign on both impurities, respectively. 
For a pure SU(2)-invariant state with vanishing impurity magnetization, $F_Q(S^{z,\pm}_{\rm II}) = 2 \pm \frac{8}{3}C_{\rm II}$. Consequently, this QFI provides complementary entanglement witnesses across the phase diagram, with $\sum_{\eta = \pm }F_Q(S^{z,\eta}_{\rm II}) = 4$, and the difference of two QFI measurements $F_Q(S^{z,+}_{\rm II}) - F_Q(S^{z,-}_{\rm II}) = 16 C_{\rm II} / 3$ provides direct access to the spin-spin correlation function. 
Similarly, we define the following generators for the closest conduction electron spins
$
    s^{z,\pm}_{\rm ee} = s^z_{{\rm e},1} \pm s^z_{{\rm e},2} 
$ and a spin-resolved density imbalance
$
   n^{\pm}_{\sigma,{\rm ee}} = n_{\sigma,{\rm e},1} \pm n_{\sigma,{\rm e},2} 
$;
the latter generator can be probed with spin-selective local-potential modulations with equal or opposite phases. Here, $n_{\sigma,{\rm e},\alpha}$ counts the number of spin-$\sigma$ electrons on the lattice site closest to the impurity on chain $\alpha$. The components of $s^{z,\pm}_{\rm ee}$ are represented in terms of electron operators as $s^{\beta}_{{\rm e},\alpha} = \frac{1}{2} \sum_{\sigma, \sigma'} c_{{\rm e},\alpha,\sigma}^\dagger (\tau^\beta)_{\sigma\sigma'} c_{{\rm e},\alpha,\sigma'}$, where $\tau^{\beta}$ is the corresponding Pauli matrix for $\beta = x,y,z$. 
Note that $n^{\pm}_{\sigma,{\rm ee}}$ can be reconstructed from independent measurements of spin and charge channels. For the total density imbalance $n^{\pm}_{{\rm ee}} = n_{{\rm e},1} \pm n_{{\rm e},2}$, where $n_{{\rm e},\alpha} = \sum_{\sigma} n_{\sigma,{\rm e},\alpha}$ has local spectral width 2.
Thus, separable states obey $F_Q(n^{\pm}_{{\rm ee}}) \leq 8$, and the maximum value is $F_Q=16$.

%%%%%%%%%%%%%%%%%%%%%%%%%%%%%%%%%%%%%%%%%%%%%%%%%%%%%%%%%
\section*{Acknowledgements} \label{sec:acknowledgements}
The authors thank Eduardo da Silva Neto, Subir Sachdev, Antoine Georges, Daniel Kaplan, Yu He, Qimiao Si, Piers Coleman, and Andrei Ruckenstein for helpful discussions, Samuele Giuli for support with NRG calculations, and Miles Stoudenmire and Uli Schollwöck for discussions on DMRG calculations.
M.C.W. acknowledges the hospitality of the Center for Computational Quantum Physics at the Flatiron Institute. The Flatiron Institute is a division of the Simons Foundation. The work of A.J.M. was supported in part by the National Science Foundation (NSF) MRSEC program through the Center for Precision-Assembled Quantum Materials (PAQM) under grant DMR-2011738. 

\section*{Contributions} \label{sec:contributions}
A.J.M. and M.C.W. defined the project. M.C.W. implemented the code, performed numerical simulations, and analyzed the data. M.C.W. wrote the manuscript with assistance from A.J.M. A.J.M. supervised the project.
%TC:endignore

%TC:ignore
\newpage
\bibliography{ref}

% REMOVE FOR DOCUMENT WITHOUT APPENDIX
% 
\newpage
\onecolumngrid
\appendix*
\newpage
\section{} 
\subsection{Experimental measurement of quantum Fisher information}
For a density matrix $\rho$ and a Hermitian generator $\mathcal{O}$, the QFI $F_Q[\rho,\mathcal{O}]$ quantifies the sensitivity of $\rho$ to the unitary transformation $\rho_\nu = e^{-i\nu \mathcal{O}} \rho \, e^{i\nu \mathcal{O}}$\cite{PhysRevA.85.022322}, where $\nu$ is the parameter encoded by the transformation. 
In the eigenbasis of $\rho=\sum_n \lambda_n \ket{n}\bra{n}$, the QFI is
\begin{equation}\label{eq:qfi_mixed}
    F_Q[\rho,\mathcal{O}]
    =
    2 \sum_{m,n}
    \frac{(\lambda_m-\lambda_n)^2}{\lambda_m+\lambda_n}
    \left|\mel{m}{\mathcal{O}}{n}\right|^2 ,
\end{equation}
only including terms with $\lambda_m+\lambda_n>0$. 
For pure states $\rho = \ket{\psi}\bra{\psi}$, which all MPS considered here are, the QFI is proportional to the operator variance ${\rm Var}(\mathcal{O})$. For mixed states, such as a thermal state $\rho_T=e^{-\beta H}/Z$ in NRG calculations, where $Z = \Tr(e^{-\beta H})$ is the thermal partition function at inverse temperature $\beta=1/T$ in units of $\hbar=k_B=1$, the QFI isolates the quantum part of the fluctuations.

Hauke et al.\cite{nphys3700} showed that the QFI can be obtained from response functions.
If the system is perturbed by a weak time-dependent field conjugate to $\mathcal{O}$
\begin{equation}
    H_{\rm pert}(t) = - h(t)\mathcal{O}\,,
\end{equation}
the retarded susceptibility is
\begin{equation}\label{eq:response}
    \chi^R_{\mathcal{O}\mathcal{O}}(t)
    =
    i\Theta(t)\,
    \ev{[\mathcal{O}(t),\mathcal{O}(0)]}_{T}.
\end{equation}
The imaginary, dissipative part $\chi''_{\mathcal{O}\mathcal{O}}(\omega,T) = \operatorname{Im}\chi^R_{\mathcal{O}\mathcal{O}}(\omega,T)$ is obtained from a Fourier transform of equation \eqref{eq:response}. 
If the experiment measures the dynamical structure factor
$S_{\mathcal{O}\mathcal{O}}(\omega,T)$, the fluctuation-dissipation theorem relates it to
$\chi''_{\mathcal{O}\mathcal{O}}(\omega,T)$ according to
\begin{equation}
    \chi''_{\mathcal{O}\mathcal{O}}(\omega,T)
    =
    \pi\left(1-e^{-\beta\omega}\right)
    S_{\mathcal{O}\mathcal{O}}(\omega,T),
     \omega>0 \,.
\end{equation}
For a non-degenerate ground state, the zero-temperature limit of equation \eqref{eq:qfi_response} reduces to
\begin{equation}
    F_Q[\ket{\psi_0},\mathcal{O}]
    =
    \frac{4}{\pi}
    \int_0^\infty \dd{\omega}\,
    \chi''_{\mathcal{O}\mathcal{O}}(\omega,0).
\end{equation}

The corresponding linear response to a generator $\mathcal{O}_{\pm}$ is obtained by applying
\begin{equation}
    \delta H_{\pm}(t)
    =
    -
    \lambda_{\pm}
    \cos(\omega t)\,
    \mathcal{O}_{\pm},
    \label{eq:ac_drive_plus_minus}
\end{equation}
and measuring the induced oscillation
\begin{equation}
    \delta
    \langle
        \mathcal{O}_{\pm}(\omega)
    \rangle
    =
    \chi_{\pm}(\omega)\,
    \lambda_{\pm}(\omega).
\end{equation}
The dissipative component $\chi_{\pm}^{\prime\prime}(\omega)$ is given by the out-of-phase response to the applied drive. Equivalently, the two channels can be reconstructed from the complete local response matrix
\begin{equation}
    \chi_{ij}(t)
    =
    i
    \Theta(t)
    \left\langle
        \left[
            \mathcal{O}_i(t),
            \mathcal{O}_j(0)
        \right]
    \right\rangle,
    \label{eq:response_matrix}
\end{equation}
with $i,j\in\{1,2\}$ through
\begin{equation}
    \chi_{\pm}(\omega)
    =
    \chi_{11}(\omega)
    +
    \chi_{22}(\omega)
    \pm
    \chi_{12}(\omega)
    \pm
    \chi_{21}(\omega).
    \label{eq:response_plus_minus}
\end{equation}
Thus, simultaneous phase-controlled driving of the two sites is not necessary. One may instead perturb each site separately, measure the local and cross responses, reconstruct the four matrix elements in equation \eqref{eq:response_matrix}, and form the symmetric and antisymmetric susceptibilities afterwards.
Because the spin-resolved occupation operators can be decomposed as
\begin{equation}
    n_{i,\uparrow}
    =
    \frac{1}{2}n_{i}
    +
    s_i^z,
    \qquad
    n_{i,\downarrow}
    =
    \frac{1}{2}n_{i}
    -
    s_i^z,
    \label{eq:spin_resolved_density_decomposition}
\end{equation}
the response of spin-selective local-potential modulations $n^{\pm}_{\sigma,{\rm ee}}$ may alternatively be reconstructed by resolving the charge ($n^{\pm}_{{\rm ee}}$) and spin ($s^{z,\pm}_{{\rm ee}}$) channels, such that
\begin{equation}
   n^{\pm}_{\uparrow,{\rm ee}}
    =
    \frac{1}{2}
    n^{\pm}_{{\rm ee}}
    +
    s^{z,\pm}_{{\rm ee}},
    \qquad
    n^{\pm}_{\downarrow,{\rm ee}}
    =
    \frac{1}{2}
    n^{\pm}_{{\rm ee}}
    -
    s^{z,\pm}_{{\rm ee}}.
    \label{eq:spin_resolved_reconstruction}
\end{equation}

\subsection{Jones--Varma critical point}
\subsubsection{Kondo temperature}
The relevant low-energy scale in the two-impurity Kondo problem for antiferromagnetic couplings $J,K>0$ is the single-impurity Kondo temperature $T_K(J)$, which determines the energy scale below which an impurity spin becomes screened by the conduction electrons. For the Kondo interaction, the weak-coupling renormalization group equation for one impurity coupled to one
conduction channel is
\begin{equation}
    \frac{dJ}{d\ell}
    =
    2\rho J^2,
    \qquad
    \ell=\ln\frac{D_0}{D},
\end{equation}
where $D_0$ is the bare bandwidth cutoff, $D$ is the running cutoff, and
$\rho$ is the local density of states per spin at the Fermi energy at the site
coupled to the impurity. Integrating this equation gives
\begin{equation}
    \frac{1}{J(D)}
    =
    \frac{1}{J}
    -
    2\rho \ln\frac{D_0}{D}.
\end{equation}
The Kondo scale is defined by the scale at which the perturbative running
coupling diverges,
\begin{equation}
    \frac{1}{J}
    -
    2\rho \ln\frac{D_0}{T_K}
    =
    0,
\end{equation}
which gives\cite{PWAnderson_1970,DONIACH1977231}
\begin{equation}
    T_K(J)
    \sim
    D_0
    \exp\left(
        -\frac{1}{2\rho J}
    \right).
\end{equation}

For the geometry considered here, each impurity is coupled to the endpoint of an
open tight-binding chain. The single-particle eigenstates of an open chain are
\begin{equation}
    \phi_n(j)
    =
    \sqrt{\frac{2}{L+1}}
    \sin(k_n j),
\end{equation}
where $k_n=\frac{\pi n}{L+1}, n = 1,\dots,L$. The dispersion is $\epsilon(k)=-2t\cos k$.  Because the impurity couples to the boundary site, the density of states entering the Kondo scale is the boundary local density of states,
\begin{equation}
    \rho_{\rm end}(\epsilon)
    =
    \int_0^\pi
    \frac{dk}{\pi}
    2\sin^2 k\,
    \delta(\epsilon+2t\cos k).
\end{equation}
Using $\left|\frac{d\epsilon}{dk}\right| = 2t\sin k$, one obtains
\begin{equation}
    \rho_{\rm end}(\epsilon)
    =
    \frac{1}{\pi t}
    \sqrt{
        1-\left(\frac{\epsilon}{2t}\right)^2
    }.
\end{equation}
At half filling, $\epsilon_F=0$ and $k_F=\pi/2$, such that $\rho_{\rm end}(0) = \frac{1}{\pi t}$. Therefore, for an impurity coupled to the endpoint of a half-filled open chain,
\begin{equation}
    T_K(J)
    \sim
    2t
    \exp\left(
        -\frac{\pi t}{2J}
    \right),
\end{equation}
where we take the natural tight-binding cutoff $D_0\sim 2t$.
The phase boundary between the Kondo-screened regime and the inter-impurity-singlet regime is controlled by the condition $K_c(J) \sim T_K(J)$.
The corresponding Kondo screening length is
\begin{equation}
    \xi_K
    \sim
    \frac{v_F}{T_K} \sim \exp\left(\frac{\pi t}{2J}\right) \,,
\end{equation}
where for the half-filled tight-binding chain $v_F = 2t$.
Therefore, finite-size DMRG calculations can only resolve the asymptotic
Kondo-screened regime when $L \gg \xi_K(J)$, or equivalently when the finite-size level spacing is small compared to the
Kondo scale, $\Delta_L \sim \frac{v_F}{L} \ll T_K(J)$. This condition becomes exponentially difficult to satisfy at small $J$, which explains why the weak-coupling phase boundary is hard to resolve numerically.

\subsubsection{Finite-size extrapolation of the critical point}
For a finite open chain, the sharp two-impurity Kondo quantum critical point is rounded by the finite-size level spacing, which acts as an infrared cutoff. Close to the critical point, the leading relevant perturbation is proportional to $K-K_c$, and the associated crossover scale behaves as
\begin{equation}
    T^\ast \sim \frac{(K-K_c)^2}{T_K}.
\end{equation}
The infrared cutoff is the finite-size level spacing
\begin{equation}
    \Delta_L \sim \frac{\pi v_F}{L+1},
\end{equation}
so the transition is rounded whenever $T^\ast \lesssim \Delta_L$. This implies a finite-size critical window
\begin{equation}
    |K-K_c| \sim \sqrt{T_K \Delta_L} \sim L^{-1/2},
\end{equation}
and motivates extrapolating pseudo-critical points according to
\begin{equation}
    K_c(L) = K_c + \frac{a}{\sqrt{L}}.
\end{equation}
Thus, for the largest system size in this study $L=500$, the level spacing is $\Delta_L \sim 0.01$, which is similar to the grid spacing of $\Delta K = 0.005$ used to calculate the data in Fig.~\ref{fig:critical_point}.

The logarithmic growth of the peak heights of local derivatives can be understood as a boundary-critical effect of the two-impurity Kondo critical point. In the two-bath geometry, tuning the inter-impurity exchange $K$ through $K_c$ couples to the leading relevant boundary operator of the impurity critical theory. The Affleck--Ludwig--Jones conformal-field-theory description\cite{PhysRevB.52.9528} identifies this perturbation with an Ising-sector boundary operator $\epsilon$ of scaling dimension $x_\epsilon=1/2$. For a local impurity or boundary observable $O$, the derivative with respect to $K$ may be written schematically as
\begin{equation}
    \partial_K \langle O\rangle
=
-\int d\tau\,
\langle O(\tau)\,(\mathbf S_1\cdot\mathbf S_2)(0)\rangle_c \,,
\end{equation}
where $\langle \dots \rangle_c$ denotes the connected correlator.
If both $O$ and $\mathbf S_1\cdot\mathbf S_2$ have nonzero overlap with the same leading boundary scaling field $\epsilon$, then at criticality
\begin{equation}
\langle \epsilon(\tau)\epsilon(0)\rangle
\sim
\frac{1}{\tau^{2x_\epsilon}}
=
\frac{1}{\tau}.
\end{equation}
In a finite chain, the infrared cutoff of the imaginary-time integral is set by the finite-size level spacing, $\tau_{\rm max}\sim L/v$, and therefore
\begin{equation}
\partial_K \langle O\rangle_{\rm peak}
\sim
A_O\int^{L/v}\frac{d\tau}{\tau}
=
A_O\log L+B_O .
\end{equation}
This argument is strongest for $C_{\rm II}=\langle \mathbf S_1\cdot\mathbf S_2\rangle$, since $C_{\rm II}=\partial_K E_0$ and hence $\partial_K C_{\rm II}=\partial_K^2 E_0$ is the direct boundary analogue of a local specific heat with respect to the tuning parameter $K$. It is also plausible for the electron--impurity correlator $C_{\rm IE}$, provided this local boundary observable has a nonzero projection onto the same leading scaling field; however, this is less rigorous because symmetry or microscopic cancellations could suppress the leading singular contribution. Thus, a fit of the form
\begin{equation}
P(L)=a\log L+b
\end{equation}
is theoretically well motivated for the peak height of $\partial_K C_{\rm II}$.

\subsection{Zero-temperature DMRG calculations}
All our calculations are performed using the ITensor library\cite{itensor} with 40 DMRG sweeps in a sector of fixed particle number. We include noise during the first sweeps to ensure convergence, and use a maximum SVD truncation error of $\varepsilon = 10^{-10}$. The MPS has total length $N_{\mathrm{sites}}=2L+2$ and open boundary conditions.

\subsubsection{Entanglement entropy}
For a given MPS approximating a quantum state $\ket{\Psi}$, we can choose any bipartition into subsystems $A$ and $B$ with two orthonormal basis sets $\{\ket{\psi_i}\}$ and $\{\ket{\psi_j}\}$ of dimensions $N_i$ and $N_j$, respectively, such that the pure state is described as
\begin{equation}\label{eq:quantumstate}
    \ket{\Psi} = \sum_{ij} \psi_{ij} \ket{\psi_i}_A \ket{\psi_j}_B \,,
\end{equation}
which defines the coefficient $\psi_{ij}$ of the $(N_i \times N_j)$-matrix $\Psi$.
This allows us to compute the reduced density matrix in system $A$ as
\begin{equation}
    \rho_A = \Tr_B(\ket{\Psi}\bra{\Psi}) = \Psi \Psi^{\dagger}\,,
\end{equation}
and similarly for $\rho_B = \Psi^{\dagger} \Psi$.\cite{dmrg_uli} We can perform a singular value decomposition (SVD) to express $\Psi$ with three new matrices
\begin{equation}
    \Psi = U S V^{\dagger} \,.
\end{equation}
Here, $S = \text{diag}(s_1, s_2, ..., s_r,0,...)$ is a matrix with singular values $s_i$ on the diagonal, where $r = \min{(N_i,N_j)}$, and $U$ and $V$ are unitary matrices.
The singular values $s_i^2 = \lambda_i$ correspond to the eigenvalues $\lambda_i$ of the Schmidt decomposition of reduced density operators for A:
\begin{equation}
    \rho_A = \sum_{i=1}^r \lambda_i \ket{\psi_i}_A \bra{\psi_i}_A \,.
\end{equation}
Therefore, the von Neumann entropy $S(\rho)$ can be calculated as
\begin{equation}
    S(\rho) = - \sum_{i=1}^{r} s_i^2 \log{s_i^2}
\end{equation}
from the singular values of the matrix $\Psi$.
Note that the entropy of the whole system is zero for pure quantum states: $S(\rho_{AB}) = 0$. The mutual information between two subsystems $A$ and $B$ is defined as
\begin{equation}\label{eq:mutual_information}
    I(A : B) = S(\rho_A) + S(\rho_B) - S(\rho_{AB}) \,.
\end{equation}
Thus, the entanglement entropy is proportional to the mutual information for our purposes.
In the nondegenerate SU(2)-invariant singlet sector, each individual impurity is maximally mixed, $S(I_1)=S(I_2)=\ln 2$. Since the full ground state is pure, the mutual informations satisfy \begin{equation} I(I_1:I_2)=2\ln 2-S(A_{\mathrm{imp}}), \qquad I(A_{\mathrm{imp}}:A_{\mathrm{el}})=2S(A_{\mathrm{imp}}), \end{equation} where $A_{\mathrm{el}}$ denotes all conduction-electron degrees of freedom. Consequently, \begin{equation}
I(I_1:I_2)+\frac{1}{2}I(A_{\mathrm{imp}}:A_{\mathrm{el}})=2\ln 2. \end{equation} This identity provides an exact identity for the transfer of total correlations from the impurity-bath sector to the impurity-impurity sector. Note that this is not a conservation law for pairwise entanglement, since the impurity mutual information contains both quantum and classical correlations.

The impurity entropy has a particularly transparent interpretation in the SU(2)-invariant singlet sector. The reduced density matrix of the two impurities can be written as 
\begin{equation} 
\rho_{\mathrm{imp}} = p_s\ket{s}\!\bra{s} + \frac{1-p_s}{3} \sum_{m=-1}^{1}\ket{t_m}\!\bra{t_m}, 
\end{equation} 
where $p_s=\frac{1}{4}-C_{\rm II}$, and $\ket{s}$ and $\bra{t}$ denote singlet and triplet states, respectively. Consequently, 
\begin{equation} 
S(A_{\mathrm{imp}}) 
= -p_s\ln p_s -(1-p_s)\ln\!\left(\frac{1-p_s}{3}\right). \end{equation} 
Thus, as described in the main text, $C_{\rm II}=0$ corresponds to a maximally mixed four-dimensional impurity sector with $S(A_{\mathrm{imp}})=2\ln2$, while $C_{\rm II}\rightarrow-3/4$ gives a pure inter-impurity singlet and $S(A_{\mathrm{imp}})\rightarrow0$. In the screened-triplet limit, $C_{\rm II}\rightarrow1/4$ and $S(A_{\mathrm{imp}})\rightarrow\ln3$.
For the same SU(2)-invariant reduced state, the pairwise entanglement between the two impurity spins is quantified by the concurrence 
\begin{equation} 
\mathcal{C}_{12} = \max\left\{0,-2C_{\rm II}-\frac{1}{2}\right\}.
\end{equation} 
The impurities are therefore pairwise entangled only when $C_{\rm II}<-1/4$. This distinguishes the growth of direct impurity--impurity entanglement from the more general left--right entanglement measured by $S(A_L)$.

\subsubsection{Convergence}
We assess the accuracy of the DMRG ground states by performing calculations at several maximum bond dimensions $\chi_{\mathrm{max}}$. The energy-variance density,
\begin{equation}
    \varepsilon(L,\chi_{\mathrm{max}})
    =
    \frac{\langle H^2\rangle-\langle H\rangle^2}{N_{\mathrm{sites}}}\,,
\end{equation}
vanishes for an exact eigenstate ($\varepsilon(L,\chi_{\mathrm{max}} \to \infty) \to 0$) and provides a global measure of the residual error of the variational MPS. 
The results presented throughout this work are obtained using the largest available bond dimension, $\chi_{\mathrm{max}}=4000$. Figs. \ref{fig:phase_diagram} and \ref{fig:qfi_ferromagnetic_transitions} contain data for $L=500$, while Fig. \ref{fig:critical_point} includes data for different lengths $L=50,100,200,500$. For all reported parameter points, we find $\varepsilon(L,\chi_{\mathrm{max}}=4000)
    \leq \mathcal{O}(10^{-7})$.

\begin{figure*}[htb!]
 \centering
 \begin{adjustbox}{center}
 %1
   \includegraphics[width=0.75\columnwidth]{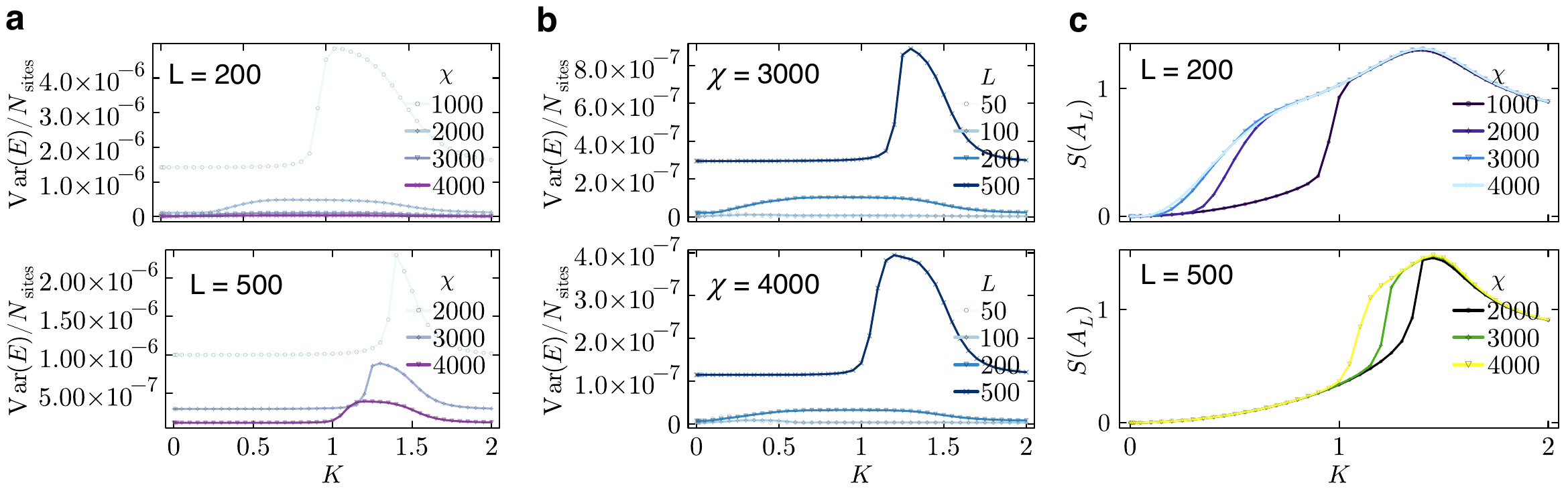} %2
 \end{adjustbox}
\caption{\textbf{Energy-variance density and bond-dimension convergence of the DMRG ground states.}
\textbf{a}, Energy variance per lattice site, $\operatorname{Var}(E)/N_{\mathrm{sites}}$, as a function of the inter-impurity exchange $K$ at fixed $J=2$, shown for different bond dimensions $\chi$ at chain lengths $L=200$ (top) and $L=500$ (bottom).
\textbf{b}, The same quantity for different chain lengths $L$ at fixed bond dimensions $\chi=3000$ (top) and $\chi=4000$ (bottom). The energy-variance density decreases systematically with increasing bond dimension. 
\textbf{c}, Entanglement entropy across the central bipartition, $S(A_L)$, for different bond dimensions at $L=200$ (top) and $L=500$ (bottom).
$S(A_L)$ does not vary significantly with bond dimension around the critical point, demonstrating convergence of the entanglement entropy, while the long-range entanglement of the diverging Kondo screening cloud for $K < K_c$ is challenging to capture accurately for the largest system with $L=500$.}
\label{fig:DMRG_convergence_energy_variance_density}
\end{figure*}

As shown in Fig. \ref{fig:DMRG_convergence_energy_variance_density}, the variance density decreases systematically with increasing bond dimension. Its largest values occur near the critical region. On the Kondo-screened side, $0<K<K_c$, the growing Kondo screening cloud generates entanglement over increasingly long distances. Thus, the bipartite entropy $S(A_L)$ is enhanced and receives appreciable contributions from a broad spectrum of Schmidt values, requiring a larger bond dimension for an accurate MPS representation. A finite bond dimension truncates the smaller Schmidt values and therefore generally underestimates the entropy. For $L=500$, the results of $S(A_L)$ (Fig. \ref{fig:DMRG_convergence_energy_variance_density}c) in the range of $K \in [1,1.3]$ may be viewed as a lower bound.

\begin{figure*}[htb!]
 \centering
 \begin{adjustbox}{center}
 %1
   \includegraphics[width=1\columnwidth]{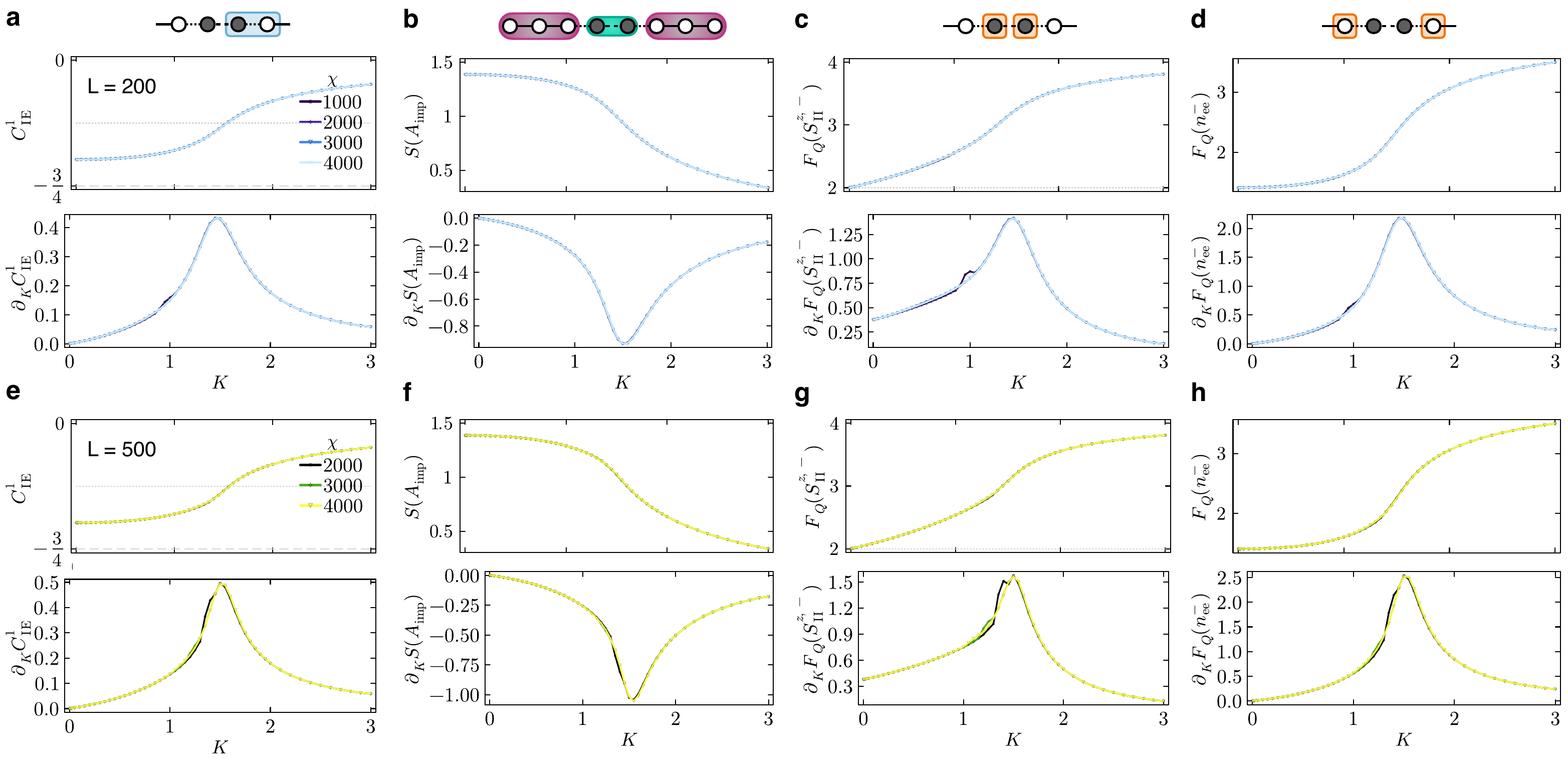} %2
 \end{adjustbox}
\caption{\textbf{Bond-dimension convergence of the DMRG observables across the critical region.}
\textbf{a--d}, Results for chain length $L=200$ and bond dimensions $\chi=1000$--$4000$.
\textbf{e--h}, Corresponding results for $L=500$ and $\chi=2000$--$4000$.
The upper panels show the local impurity-electron correlator $C_{\mathrm{IE}}^{1}$ (\textbf{a,e}), the impurity entropy $S(A_{\mathrm{imp}})$ (\textbf{b,f}), the impurity-spin QFI $F_Q(S_{\mathrm{II}}^{z,-})$ (\textbf{c,g}) and the local electron-density QFI $F_Q(n_{ee}^{-})$ (\textbf{d,h}) as functions of the inter-impurity exchange $K$, at fixed $J=2$.
The lower panels show the corresponding derivatives with respect to $K$.
The near-collapse of the curves demonstrates convergence with bond dimension. In particular, the deviation between the two largest bond dimension calculations is minimal even at the critical point.}
\label{fig:DMRG_convergence_bond_dim_extrapolation}
\end{figure*}

Figure~\ref{fig:DMRG_convergence_bond_dim_extrapolation} directly compares the correlators, entanglement entropies and QFIs presented in the main text for several bond dimensions. For $L=200$ (top) and $L=500$ (bottom), the observables themselves are nearly indistinguishable at the two largest bond dimensions.
Their derivatives with respect to $K$, which are more sensitive to small numerical differences, likewise show only minor residual deviations near the critical point, indicating convergence with respect to bond dimension.
In this work, we only report the results at the largest available bond dimension rather than extrapolating each observable to $\chi_{\mathrm{max}}\rightarrow\infty$. Although the energy variance is a useful convergence diagnostic, there is no universal requirement that a local observable be linear in the variance density. In particular, close to the critical region, different observables such as spin-spin correlations, entanglement entropies, and variances of collective operators can exhibit nonlinear and non-monotonic corrections with bond dimension. 
Extrapolating derivatives is even less robust because differentiation amplifies small differences between independently optimized MPS calculations.
We also retain the results at each finite length instead of extrapolating the full curves directly to $L\rightarrow\infty$ because the position and width of the critical features evolve with $L$, and values evaluated at fixed $K$ do not, in general, follow a simple extrapolation in $L$. Consequently, thermodynamic trends are inferred from the systematic comparison between the different lengths.

\subsubsection{QFI across the zero-temperature phase diagram}
\begin{figure*}[htb!]
 \centering
 \begin{adjustbox}{center}
 %1
   \includegraphics[width=1\columnwidth]{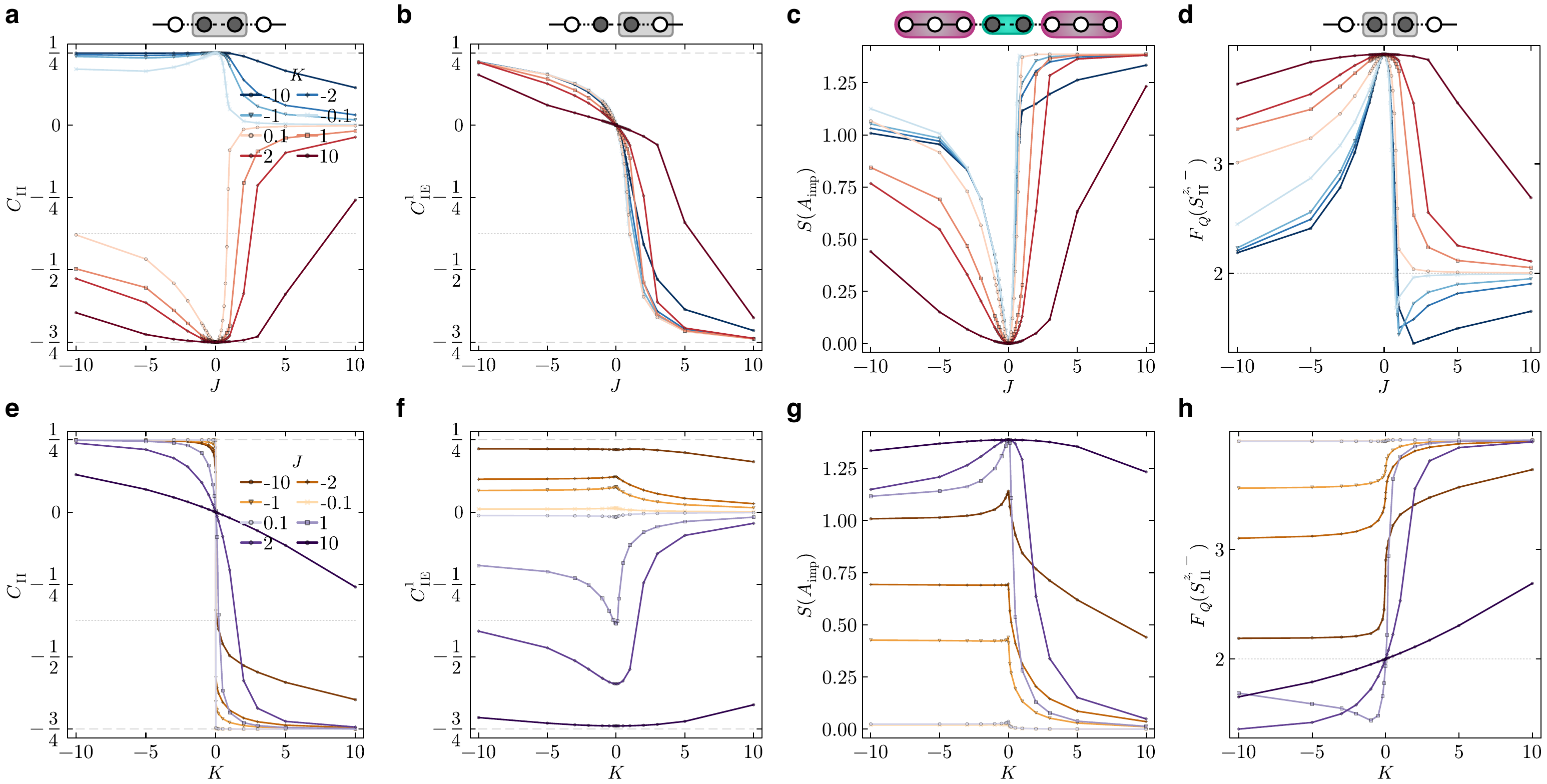}
 \end{adjustbox}
 \caption{\textbf{Linecuts of correlators, entanglement, and QFI across the phase diagram}. Observables as a function of $J$ ($K$) in top (bottom) row, corresponding to horizontal (vertical) linecuts of data from Fig. 2. \textbf{a}, Impurity-impurity correlator for different $K$ shows singlet (triplet) formation $K>0$ in blue ($K<0$ in red) and $\abs{J} \rightarrow 0$. For $\abs{J}/\abs{K} \gg 1$, the impurities become uncorrelated. For $J,K<0$, impurities are locked into a triplet state. For $J=0$, the dimer limit is recovered, with $C_{\rm II} = -3/4$ ($C_{\rm II} = 1/4$) for $K>0$ ($K<0$). \textbf{b}, Electron-impurity correlator shows AFM (FM) screening for $J>0$ ($J<0$), independent of $K$. All curves intersect at $J=0$ where $C_{\rm IE} = 0$. \textbf{c}, Entanglement of impurities with their conduction electrons is zero at $J=0$ and increases with increasing $\abs{J}$ up to saturation at $S(A_{\rm imp}) = 2 \ln{2}$. At fixed $\abs{J}$, the entropy is increasing for decreasing $\abs{K}$, although less pronounced for $K<0$ compared to $K>0$. \textbf{d}, QFI of staggered impurity spin operators witnesses bipartite entanglement (dotted line denotes threshold $F_Q = 2$) for all parameters except in quadrant $K<0, J>0$. The QFI is similar to $-S(A_{\rm imp})$, with a maximum when $S(A_{\rm imp}) = 0$. \textbf{e}, For $K/\abs{J}\gg1$, the impurities are forming a singlet ($C_{\rm II} = -3/4$); for $K < 0$, they form a triplet ($C_{\rm II} = 1/4$). Correlations to the bath become more pronounced for $\abs{K} \rightarrow 0$. \textbf{f}, Increasing $K$ leads to decreasing AFM screening and FM polarization for $J>0$ and $J<0$, respectively. Upon increasing $K>0$, the impurity-bath correlations reduce; similarly for $K<0$, but on a different scale of $K/J$ in the effective two-channel spin-1 limit. \textbf{g}, For each value of $J$, the impurity-bath entanglement is maximal at $K=0$. For $K,J<0$, the entanglement saturates to a non-zero value; while for $K/\abs{J} \gg 1$, it decreases to zero. \textbf{h}, $F_Q(S^{z,-}_{II})$ certifies entanglement except for $K<0$ and $J/\abs{K} > 1$. }
 \label{fig:figureSI_linecuts}
\end{figure*}

\begin{figure*}[htb!]
 \centering
 \begin{adjustbox}{center}
 %1
   \includegraphics[width=1\columnwidth]{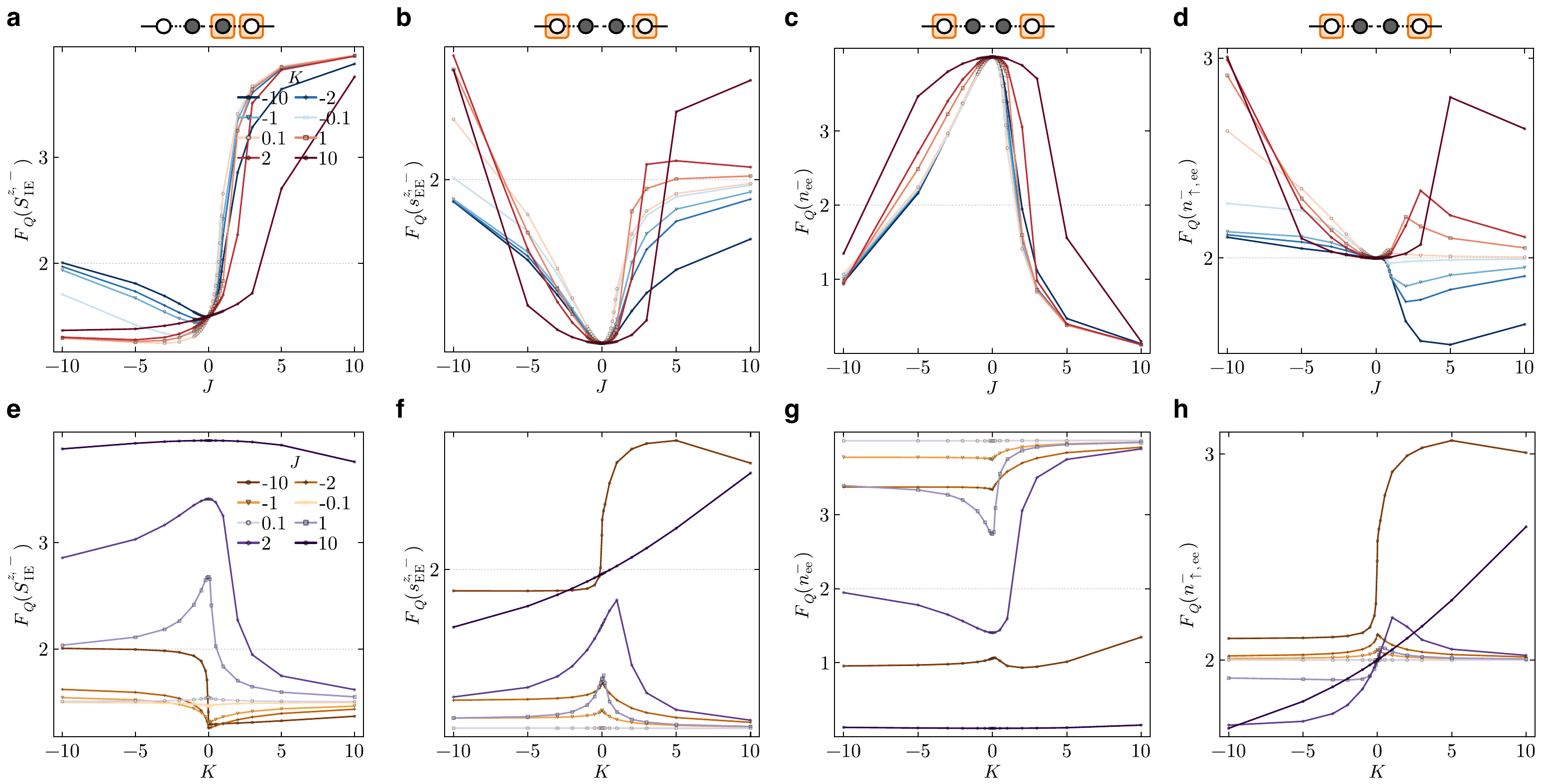}
 \end{adjustbox}
    \caption{\textbf{Linecuts of electronic quantum Fisher information.} QFI as a function of $J$ for different fixed values of $K$ (top row) and as a function of $K$ for different fixed values of $J$ (bottom row), corresponding to horizontal and vertical linecuts of the phase diagrams in Fig.~2, respectively. \textbf{a,e}, $F_Q(S^{z,-}_{\rm IE})$ of the staggered impurity--electron spin. \textbf{b,f}, $F_Q(s^{z,-}_{\rm ee})$ of the spin imbalance between the electron sites adjacent to the impurities. \textbf{c,g}, $F_Q(n^{-}_{\rm ee})$ of the corresponding charge imbalance. \textbf{d,h}, $F_Q(n^{-}_{\uparrow,\rm ee})$ of the spin-resolved charge imbalance.}
 \label{fig:figureSI_linecuts2}
\end{figure*}

The linecuts in Fig. \ref{fig:figureSI_linecuts} make explicit how the redistribution of entanglement across the phase diagram is reflected in spin-spin correlations, which can be detected through the impurity-spin QFI. For fixed inter-impurity exchange $K$, the impurity correlator approaches the isolated-dimer limits $C_{\rm II}\rightarrow-3/4$ for $K>0$ and $C_{\rm II}\rightarrow1/4$ for $K<0$ as $J\rightarrow0$, see panel a, whereas increasing $\abs{J}$ enhances the local impurity--electron correlations and weakens the inter-impurity locking. The sign of $C_{\rm IE}$ distinguishes antiferromagnetic Kondo screening for $J>0$ from ferromagnetic polarization for $J<0$, see Fig. \ref{fig:figureSI_linecuts}a, while the impurity entropy $S(A_{\rm imp})$ quantifies the corresponding growth of impurity--bath entanglement away from the decoupled line $J=0$, as shown in panel c. The complementary cuts at fixed $J$ in Fig. \ref{fig:figureSI_linecuts}e-g show the transfer of correlations from the impurity--bath sector to an inter-impurity singlet as $K$ is increased, and to a triplet-correlated regime for sufficiently negative $K$. Note that for small $J=0.1$, the Kondo temperature is $T_K \sim 10^{-7}$, such that the behavior at $J \rightarrow 0$ is not resolved.

The electronic QFI linecuts in Fig.~\ref{fig:figureSI_linecuts2} provide additional local signatures of the same redistribution of correlations identified in Fig. \ref{fig:figureSI_linecuts}. The cuts of $F_Q(S^{z,-}_{\rm IE})$ and $F_Q(s^{z,-}_{\rm ee})$ show how impurity--electron and nearby electronic-spin fluctuations evolve between the Kondo-screened, inter-impurity-singlet, and triplet-correlated regimes, while $F_Q(n^{-}_{\rm ee})$ and $F_Q(n^{-}_{\uparrow,\rm ee})$ demonstrate that this reorganization is also encoded in local charge fluctuations.

\begin{figure*}[htb!]
 \centering
 \begin{adjustbox}{center}
   \includegraphics[width=1\columnwidth]{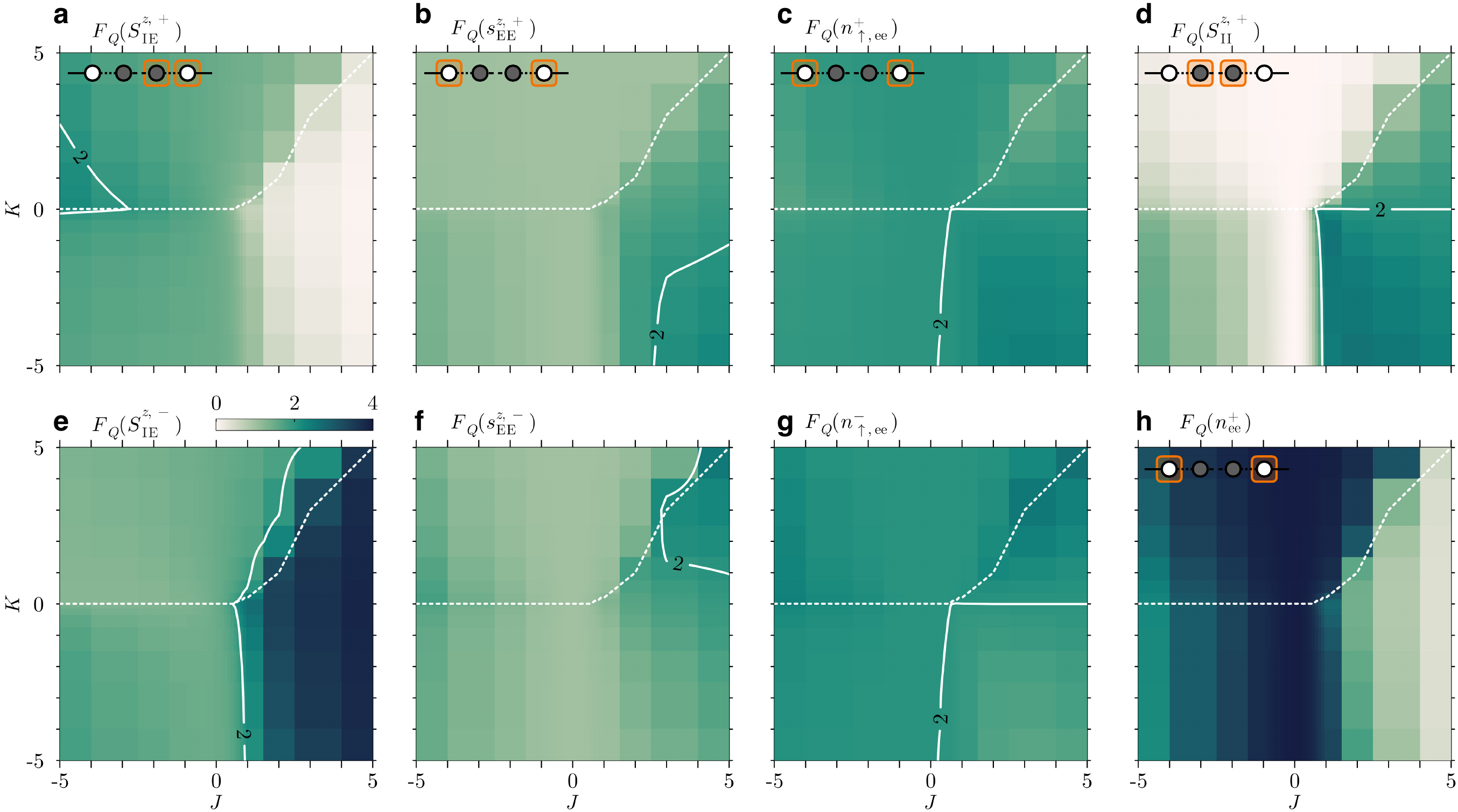}
 \end{adjustbox}
\caption{\textbf{Additional ground-state quantum Fisher information across the phase diagram.}
\textbf{a,e}, QFI $F_Q(S^{z,+}_{\rm IE})$ and $F_Q(S^{z,-}_{\rm IE})$, respectively, for the symmetric and antisymmetric spin combinations of an impurity and its nearest electron site.
\textbf{b,f}, QFI $F_Q(s^{z,+}_{\rm EE})$ and $F_Q(s^{z,-}_{\rm EE})$, respectively, for the symmetric and antisymmetric spin combinations of the two nearest electron sites.
\textbf{c,g}, QFI $F_Q(n^{+}_{\uparrow,\rm ee})$ and $F_Q(n^{-}_{\uparrow,\rm ee})$, respectively, for the symmetric and antisymmetric spin-resolved density combinations of the nearest electron sites.
\textbf{d}, QFI $F_Q(S^{z,+}_{\rm II})$ for the symmetric impurity-spin operator.
\textbf{h}, QFI $F_Q(n^{+}_{\rm ee})$ for the symmetric total-density combination of the nearest electron sites. Dashed white lines indicate the contour where $C_{\rm II} = -3/8$, while solid white contours mark the two-particle separability bound $F_Q=2$ (not shown in \textbf{h}).}
 \label{fig:figureSI_phase_diagram}
\end{figure*}

Fig. \ref{fig:figureSI_phase_diagram} presents an overview of the QFI across the phase diagram: $F_Q(S^{z,-}_{\rm II})$ is enhanced in singlet-correlated regions, whereas $F_Q(S^{z,+}_{\rm II})$ is enhanced in triplet-correlated regions, such that the two channels provide complementary entanglement witnesses; the phase diagram of the symmetric impurity channel is shown in Fig.~\ref{fig:figureSI_phase_diagram}(d). The corresponding impurity--electron fluctuations in Fig. \ref{fig:figureSI_phase_diagram}a and e distinguish the two types of local impurity--bath locking: $F_Q(S^{z,+}_{\rm IE})$ is enhanced predominantly for ferromagnetic impurity--electron coupling, while $F_Q(S^{z,-}_{\rm IE})$ becomes large in the antiferromagnetic Kondo-screened regime. The same correlation pattern is transferred to the conduction electrons closest to the impurities. In Fig. \ref{fig:figureSI_phase_diagram}b and f, the antisymmetric and symmetric electronic-spin operators $s^{z,-}_{\rm ee}$ and $s^{z,+}_{\rm ee}$ retain signatures of the singlet- and triplet-correlated regions, respectively, although they do not provide equally broad entanglement certification throughout the phase diagram. By contrast, the spin-resolved density operators $n^{+}_{\uparrow,\rm ee}$ and $n^{-}_{\uparrow,\rm ee}$ in panels (c) and (g) of Fig. \ref{fig:figureSI_phase_diagram}, which combine local charge and spin fluctuations, exceed the two-site separability bound in complementary regions of parameter space. Fig. \ref{fig:figureSI_phase_diagram}h shows the symmetric spin-summed density fluctuation $F_Q(n^{+}_{\rm ee})$, which is equivalent to $F_Q(n^{-}_{\rm ee})$. It also carries a pronounced imprint of the correlation structure, with its strongest enhancement in the ferromagnetic-coupling regime. These results demonstrate that the rearrangement of quantum correlations is not confined to the impurity sector, but can also be detected through spatially local spin and charge fluctuations of the surrounding conduction electrons.

\subsubsection{QFI across the critical point}
\begin{figure*}[htb!]
 \centering
 \begin{adjustbox}{center}
 %1
   \includegraphics[width=1\columnwidth]{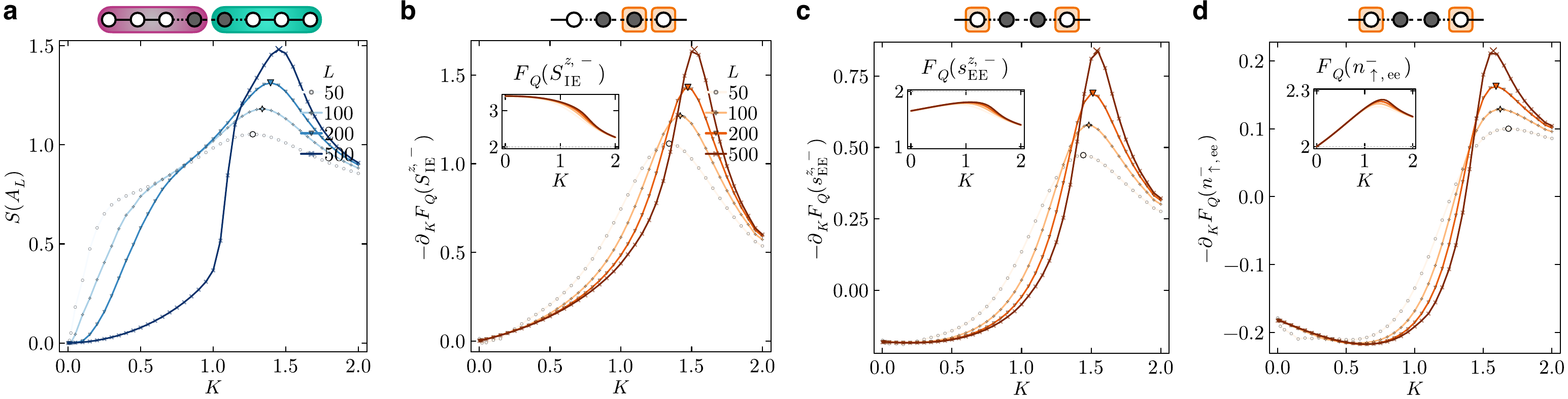}
 \end{adjustbox}
 \caption{\textbf{Additional linecuts across the impurity quantum critical point.} Observables as a function of $K$ for $J=2$ and different electron-chain lengths $L$. \textbf{a}, Entanglement entropy $S(A_L)$ between the two chains. \textbf{b--d}, Negative derivatives of the QFI for the staggered impurity--electron spin, $-\partial_K F_Q(S^{z,-}_{\rm IE})$; the spin imbalance of the adjacent electron sites, $-\partial_K F_Q(s^{z,-}_{\rm ee})$; and their spin-resolved charge imbalance, $-\partial_K F_Q(n^{-}_{\uparrow,{\rm ee}})$, respectively. The insets show the corresponding undifferentiated QFI. All observables exhibit a size-dependent feature that follows the critical peak toward the quantum critical point.}
 \label{fig:figureSI_QFI}
\end{figure*}

Fig. \ref{fig:figureSI_QFI} contains additional data for the linecuts for $J=2$ as a function of $K$ presented in fig. \ref{fig:critical_point}. Fig. \ref{fig:figureSI_QFI}(a) contains the central cut entropy $S(A_L)$, which is zero at $K=0$ when the impurities are uncorrelated and becomes maximal around, but not exactly at $K_c$. Panel (b) contains the derivative of the antisymmetric impurity-electron generator, $\partial_K F_Q(S^{z,-}_{\rm IE})$, which follows the impurity-electron spin correlator and hence becomes maximal at $K_c$. Measurement of the generator $S^{z,-}_{\rm IE} = S^z_1 - s^z_{e,1}$ witnesses entanglement because $F_Q(S^{z,-}_{\rm IE}) > 2$ for all values of $K>0$, see inset. The electron-spin QFI $F_Q(s^{z,-}_{ee})$ is presented in fig. \ref{fig:figureSI_QFI}(c). Similar to the spin-resolved density imbalance (panel d), it follows the critical point, but only the latter QFI also verifies entanglement, see insets. Together, these data show how different QFIs can be used to reconstruct spin-spin correlations and track the redistribution of entanglement around the critical point.

\newpage
\subsubsection{Detection of the critical point}\label{sec:kondoscreeninglength}
\begin{figure*}[htb!]
 \centering
 \begin{adjustbox}{center}
 %1
   \includegraphics[width=0.6\columnwidth]{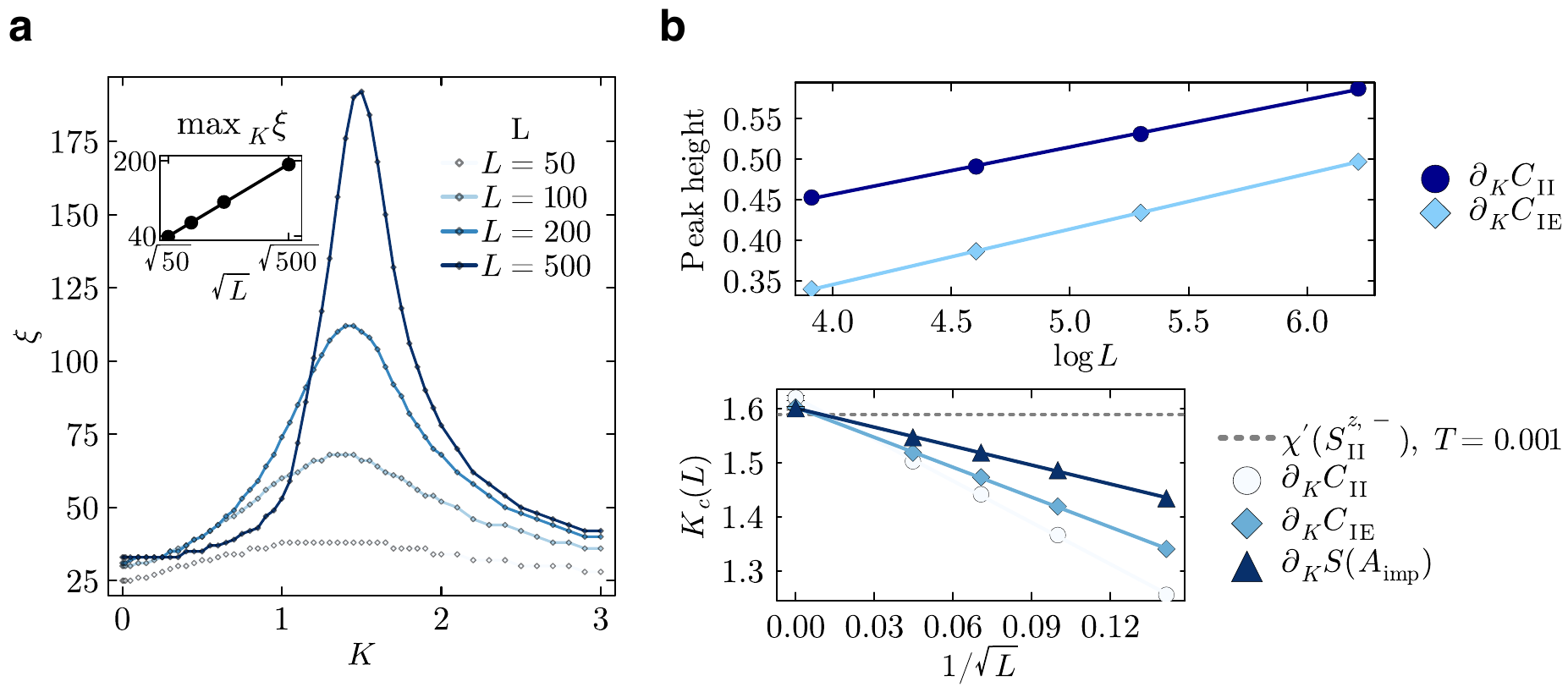}
 \end{adjustbox}
\caption{\textbf{Finite-size signatures of the Jones--Varma critical point.}
\textbf{a}, Correlation-weight screening length $\xi$ obtained from the impurity--electron spin correlator for different $L$, showing a pronounced finite-size maximum in the competition regime between Kondo screening and inter-impurity singlet formation. Inset shows scaling of peak height with $\sqrt{L}$.
\textbf{b}, Top: Peak height versus $\log{L}$ for $\partial_K C_{\mathrm{II}}$ and $\partial_K C_{\rm IE}$. Bottom: Peak positions versus $1/\sqrt{L}$ for $\partial_K C_{\mathrm{II}}$, and $\partial_K C_{\rm IE}$, $\partial_K S(A_{\rm imp})$. Dotted gray line corresponds to the peak of static susceptibility of $T=0.001$ NRG calculation. Error bars correspond to the standard error of the $L\to\infty$ intercept obtained from an unweighted linear least-squares fit.}
\label{fig:critical_point_detection}
\end{figure*}

To characterize the spatial extent and sign of the Kondo screening cloud\cite{PhysRevB.80.205114}, we define the total impurity--electron screening correlator
\begin{equation}
    C^\alpha_{\rm tot}
    =
    \sum_{d=1}^{L} C^\alpha_{\rm IE}(d) \,,
\end{equation}
where $C^{\alpha}_{\rm IE}(d) = \langle \mathbf{S}_{\alpha} \cdot \mathbf{s}_{d,\alpha} \rangle$ is the correlator between impurity and electron on chain $\alpha=1,2$ separated by a distance $d=1,...,L$. Here, $C^\alpha_{\rm tot}<0$ signals an AFM screening cloud, while $C^\alpha_{\rm tot}>0$ indicates FM polarization of the conduction electrons around the impurity. We introduce the normalized accumulated correlation weight
\begin{equation}
    P_\alpha(r)
    =
    \frac{
    \sum_{d=1}^{r} \left| C^\alpha_{\rm IE}(d) \right|
    }{
    \sum_{d=1}^{L} \left| C^\alpha_{\rm IE}(d) \right|
    }
\end{equation}
with $r=1,\dots,L$.
This quantity measures the distance over which a fraction $p$ of the absolute impurity-electron correlation weight is accumulated, independent of the sign structure of the correlations. The corresponding correlation-weight screening length is defined as
\begin{equation}
    \xi^\alpha(p)
    =
    \min
    \left\{
    r \in \{1,\dots,L\}
    :
    P_\alpha(r) \ge p
    \right\}.
\end{equation}
In practice, we use $p=0.98$ and denote the corresponding length by $\xi^\alpha \equiv\xi^\alpha(p = 0.98)$. We verified that $\xi^1 = \xi^2$ and omit the chain index $\alpha$. Note that this length is not equivalent to the physical Kondo length $\xi_K \sim v_F / T_K$.

Fig. \ref{fig:critical_point_detection}a contains the correlation-weight screening length $\xi$ extracted from the impurity--electron spin correlator as a function of $K$ for $J=2$. We observe a pronounced maximum whose position drifts systematically with system size, reflecting the growth and eventual cutoff of the screening cloud near criticality, see Fig. \ref{fig:critical_point_detection}a. The inset shows the maximum correlation-weight screening length $\max_K{\xi}$ at the peak which scales with $\sqrt{L}$.
To extract the thermodynamic critical point, we compare the finite-size peak positions of $\partial_K C_{\rm II}$, $\partial_K C_{\rm IE}$, and $\partial_K S(A_{\rm imp})$ as functions of $1/\sqrt{L}$ in Fig. \ref{fig:critical_point_detection}b. The extrapolated intercept for $\partial_K C_{\mathrm{II}}$ yields $K_c = 1.620 \pm 0.005$, which is larger than the single-impurity Kondo scale $T_K\sim 0.91$. Panel (b) also includes the peak of the static susceptibility $\chi'(S^{z,-}_{\mathrm{II}})$ at $K_c = 1.59 \pm 0.03$ for $T=0.001$, which is obtained from NRG.
The error bars denote the standard error of the $L\to\infty$ intercept obtained from an unweighted linear least-squares fit of $K_c(L)$ versus $1/\sqrt{L}$, with the residual variance estimated from the fit residuals. 

The comparison between the lowest-temperature NRG results and the finite-size DMRG data must account for the different infrared cutoffs of the two methods. Near the Jones--Varma critical point, the crossover scale behaves as $T^{*}\sim (K-K_{c})^{2}/T_{K}$, so a finite temperature rounds the transition over a coupling window $\delta K_{T}\sim\sqrt{T T_{K}}$. For $T=0.001$ and $T_{K}\simeq0.91$, this gives $\delta K_{T}\simeq0.03$, consistent with the observed displacement of the lowest-temperature NRG response maximum from the extrapolated zero-temperature value. In DMRG, the corresponding infrared cutoff is the finite-size level spacing, which for $L=500$ is $\Delta_{L}\simeq0.0125$, producing a broader finite-size critical window $\delta K_{L}\sim\sqrt{T_{K}\Delta_{L}}\simeq0.11$. 

\subsubsection{QFI across the ferromagnetic sector boundaries}
% Figure 4: two other quantum phase transitions
\begin{figure*}[htb!]
 \centering
 \begin{adjustbox}{center}
   \includegraphics[width=0.5\textwidth]{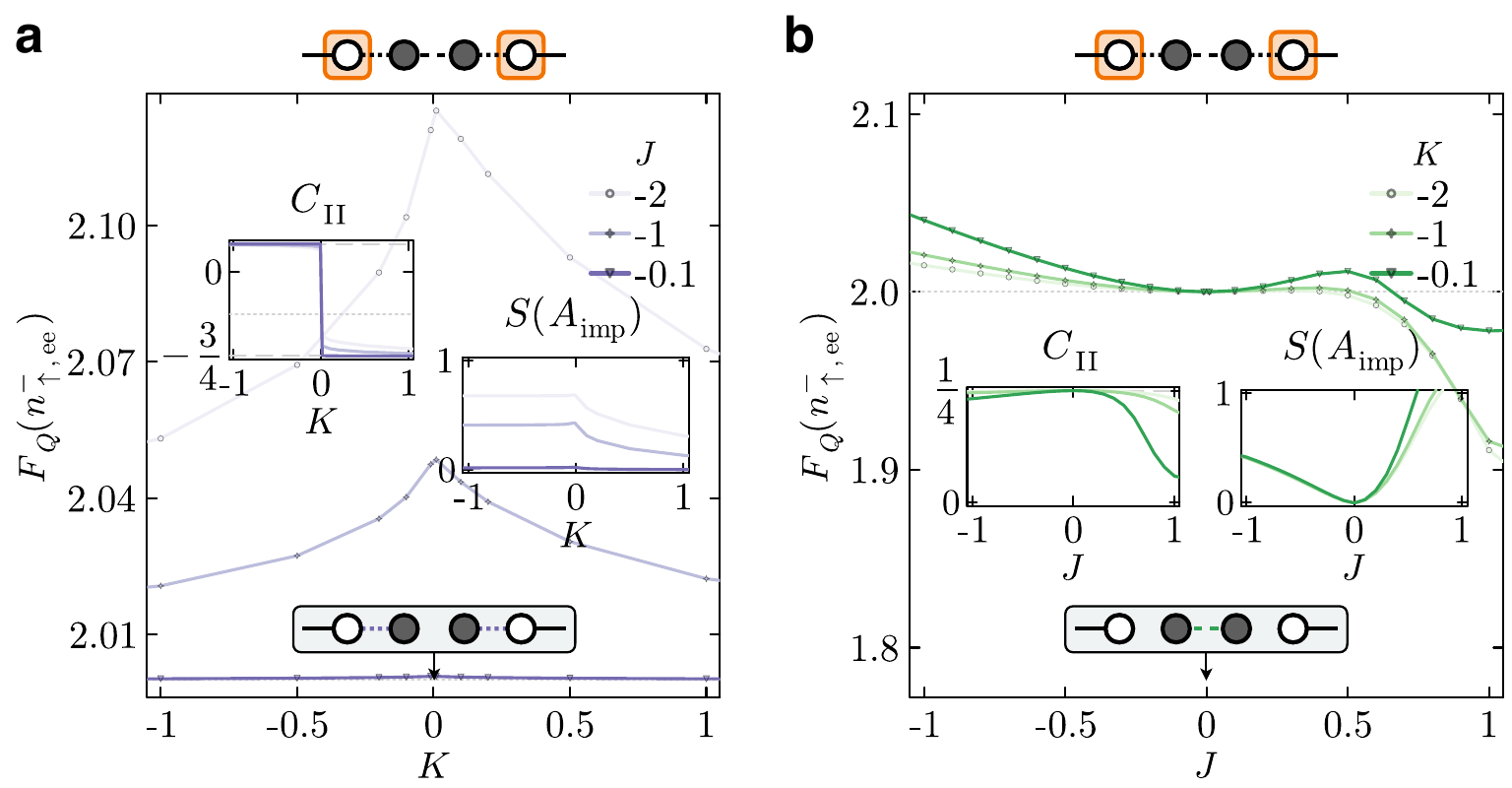} %1
 \end{adjustbox}
\caption{\textbf{Local electronic QFI resolves the ferromagnetic-sector boundaries.}
QFI $F_Q(n^{-}_{\uparrow,{\rm ee}})$ of the spin-resolved density imbalance between the electron sites adjacent to the two impurities. The horizontal dashed line marks the separability bound $F_Q=2$.
\textbf{a}, Fixed ferromagnetic impurity--electron couplings $J<0$ and varying inter-impurity exchange $K$. The QFI peaks at the level crossing of residual moments $K=0$. Insets show the corresponding impurity correlator
$C_{\mathrm{II}}$ and impurity entropy $S(A_{\mathrm{imp}})$; their sharp jump and peak, respectively, reflect the competition between inter-impurity locking and local impurity--bath dressing.
\textbf{b}, Fixed ferromagnetic inter-impurity exchanges $K<0$ and varying impurity--electron coupling $J$. The QFI develops a local minimum at the decoupled point $J=0$, where $\partial_JF_Q=0$. The insets recover the exact limits
$C_{\mathrm{II}}=1/4$
and
$S(A_{\mathrm{imp}})=0$
of an impurity triplet factorized from the electron baths.}
\label{fig:qfi_ferromagnetic_transitions}
\end{figure*}

Fig.~\ref{fig:qfi_ferromagnetic_transitions} shows the two phase boundaries connected to the decoupled axes of the two-impurity Kondo model. 
For fixed $J<0$, tuning $K$ through zero produces a singlet-triplet level crossing of residual local moments, see Fig.~\ref{fig:qfi_ferromagnetic_transitions}a, as revealed by $C_{\rm II}$, which distinguishes triplet and singlet impurity locking. The entropy $S(A_{\rm imp})$ is approximately constant for $K<0$ with a maximum around $K=0$ when the impurities are in a FM polarized triplet state, but decreases with increasing inter-impurity coupling $K>0$. 
This boundary is sharp at $T=0$, but it does not generate a divergent Kondo screening length or a non-Fermi-liquid criticality. 
The electronic density QFI detects the same rearrangement without measuring the impurity spins directly. The cusp-like enhancement around $K=0$ should be read as a local level-crossing signature.
The spin-spin correlator $C_{\rm II}$ becomes smoother with increasing $J<0$, reflecting the dependence on the ratio of $\abs{J}/K$; it follows the Hellmann-Feynman relation $\partial_K E=C_{\mathrm{II}}$ and is continuously connected to the exactly solvable dimer limits $C_{\mathrm{II}}=1/4$ for $K<0$ and $C_{\mathrm{II}}=-3/4$ for $K>0$ at $J=0$.

For fixed ferromagnetic inter-impurity exchange $K<0$, tuning $J$ through zero instead changes how the impurity triplet is coupled to the baths, see Fig.~\ref{fig:qfi_ferromagnetic_transitions}b. At $J=0$ the triplet factorizes from the conduction electrons, giving $C_{\rm II}=1/4$, $C_{\rm IE}=0$, $S(A_{\rm imp})=0$, and a local minimum of the electronic QFI. Moving to $J>0$ turns on AFM screening of the effective spin-one impurity, whereas $J<0$ flows to a FM local-moment regime with only marginal impurity-bath dressing. As shown in Fig. \ref{fig:phase_diagram}, the symmetric and antisymmetric operators acts as complementary entanglement witnesses. 
The QFI therefore witnesses entanglement between the left and right impurity–bath subsystems wherever $F_Q(n^{-}_{\uparrow,{\rm ee}})> 2$, while its extrema locate the two distinct boundaries. At $J=0$, is reaches but does not exceed the separability bound.

\begin{figure*}[htb!]
 \centering
 \begin{adjustbox}{center}
 %1
   \includegraphics[width=0.5\columnwidth]{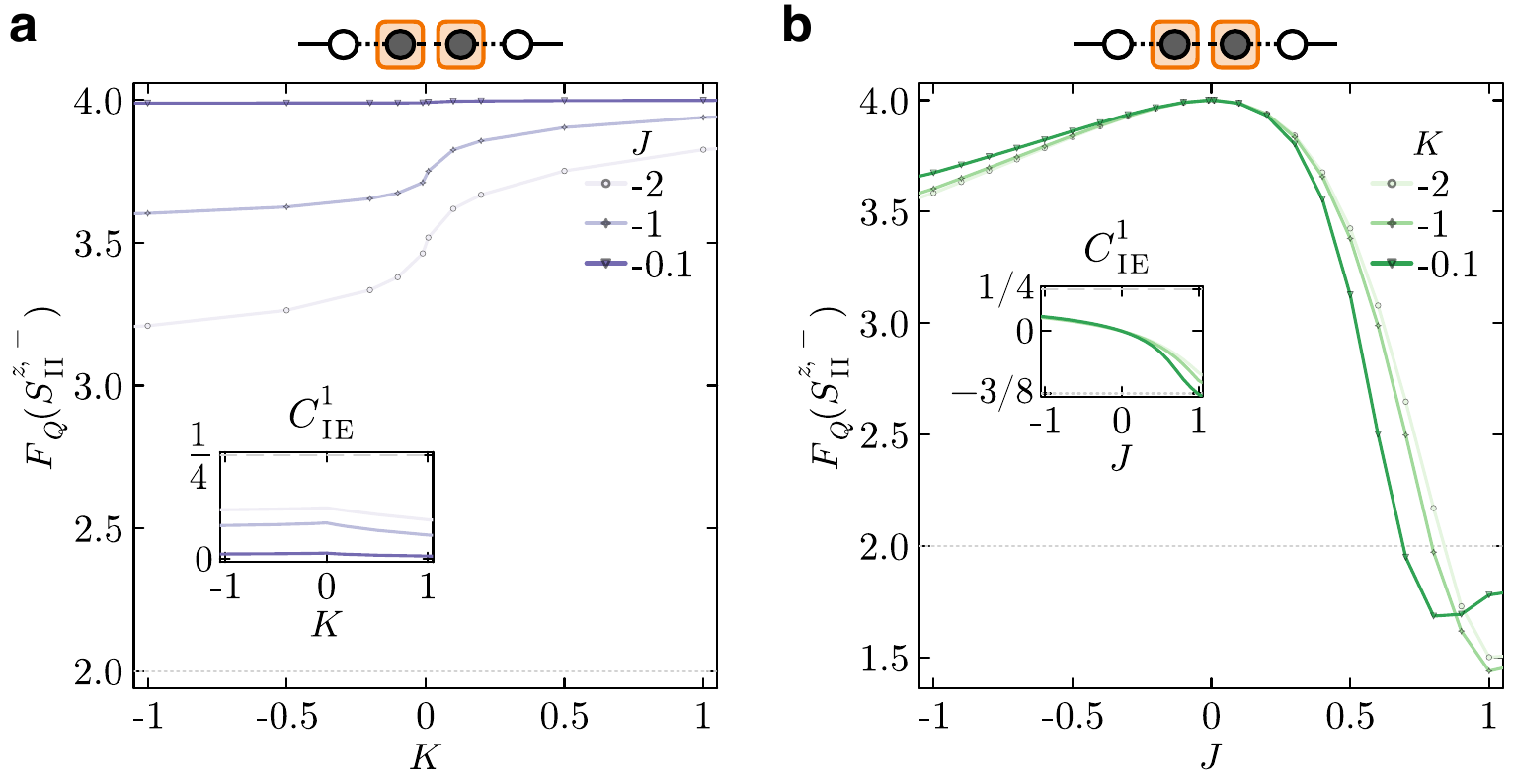}
 \end{adjustbox}
 \caption{\textbf{Additional QFI linecuts across the ferromagnetic-sector boundaries.} Impurity-spin QFI $F_Q(S^{z,-}_{\rm II})$ across the decoupled limits. \textbf{a}, As a function of $K$ across $K=0$ for different fixed $J<0$, the QFI exhibits an inflection point associated with the level crossing of the residual impurity moments. \textbf{b}, As a function of $J$ across $J=0$ for different fixed $K<0$, the QFI becomes maximal at the decoupled-impurity limit. Insets show the corresponding impurity--electron correlator $C^{1}_{\rm IE}$. The impurity-spin QFI follows the inter-impurity correlations and remains above the separability bound throughout around the transition; when $F_Q(S^{z,-}_{\rm II}) < 2$, the symmetric QFI $F_Q(S^{z,+}_{\rm II})$ can be used to witness entanglement.}
 \label{fig:figureSI_QFI_FM_sectors}
\end{figure*}

Fig. \ref{fig:figureSI_QFI_FM_sectors}(a) and (b) contain additional data of $F_Q(S^{z,-}_{\rm II})$ for the decoupled single-impurity Kondo problems across $K=0$ and for the ferromagnetic boundaries of decoupled impurities across $J=0$, respectively, corresponding to the linecuts in Fig. \ref{fig:qfi_ferromagnetic_transitions}. The impurity spin-spin QFI follows the spin-spin correlator and has an inflection point across $K=0$ or becomes maximal across $J=0$. Across the transition, this QFI certifies entanglement; away from it, $F_Q(S^{z,+}_{\rm II})$ acts as a complementary entanglement witness to $F_Q(S^{z,-}_{\rm II})$.

\subsection{Finite-temperature NRG calculations}
\subsubsection{Hamiltonian}
Finite-temperature observables are computed using full-density-matrix numerical renormalization group (FDM--NRG) calculations with the NRG Ljubljana/TRIQS implementation\cite{PhysRevB.79.085106}. The Hamiltonian is given by
\begin{equation}
\begin{aligned}
H_{\mathrm{NRG}}^{(\Lambda,z)}
={}&
\sum_{\alpha=1}^{2}\sum_{n=0}^{\infty}\sum_{\sigma}
\left[
\epsilon_{n}^{(\Lambda,z)}
f_{n\alpha\sigma}^{\dagger}f_{n\alpha\sigma}
+
t_{n}^{(\Lambda,z)}
\left(
f_{n\alpha\sigma}^{\dagger}f_{n+1,\alpha,\sigma}
+\mathrm{h.c.}
\right)
\right]
\\
&+
\sum_{\alpha=1}^{2}
\left[
U_{\mathrm{aux}}n_{\alpha\uparrow}n_{\alpha\downarrow}
-\frac{U_{\mathrm{aux}}}{2}
\left(n_{\alpha\uparrow}+n_{\alpha\downarrow}\right)
\right]
+
J\sum_{\alpha=1}^{2}
\mathbf S_{\alpha}\cdot\mathbf s_{0,\alpha}
+
K\,\mathbf S_{1}\cdot\mathbf S_{2},
\end{aligned}
\end{equation}
where $f_{n\alpha\sigma}$ denotes Wilson-chain site $n$ of bath $\alpha$, $\mathbf s_{0,\alpha}$ is the spin of the zeroth Wilson site, and the coefficients $\epsilon_{n}^{(\Lambda,z)}$ and $t_{n}^{(\Lambda,z)}$ are obtained by logarithmically discretizing the endpoint density of states of the semi-infinite tight-binding chain,
\begin{equation}
\rho_{\mathrm{end}}(\epsilon)
=
\frac{1}{\pi t}
\sqrt{1-\left(\frac{\epsilon}{2t}\right)^{2}}
\,\Theta(2t-|\epsilon|).
\end{equation}
The two independent non-interacting conduction baths are logarithmically discretized with parameter $\Lambda$ and mapped to two Wilson chains, which are diagonalized iteratively, retaining the lowest $N_{\rm keep}$ many-body multiplets below the rescaled energy cutoff $E_{\rm keep}$. 
Particle--hole symmetry gives $\epsilon_{n}^{(\Lambda,z)}=0$, while the Wilson-chain hoppings decrease asymptotically as $t_{n}^{(\Lambda,z)}\propto \Lambda^{-n/2}$.
The empty and doubly occupied auxiliary states are therefore separated from the singly occupied spin doublet by an energy of order $U_{\rm aux}/2$, so that the auxiliary orbitals act as frozen spin-$1/2$ degrees of freedom. We use $U_{\mathrm{aux}}=100t$, so for $J=2t$ one has $(J/U_{\mathrm{aux}})^{2}=4\times10^{-4}$. Even this small estimate overstates the error, because the auxiliary-orbital charge is exactly conserved and there is no hybridization that could generate virtual transitions into the empty or doubly occupied sectors; these states are additionally separated from the singly occupied spin doublet by $U_{\mathrm{aux}}/2=50t$, making their thermal population exponentially negligible on all temperatures considered.

Thermal expectation values are evaluated from the thermal density matrix $\rho_T = \frac{e^{-\beta H}}{Z}$ as $\langle A\rangle_T = \mathrm{Tr}(\rho_T A)$ using the complete discarded-state basis of the Wilson chain, where $Z = \mathrm{Tr}\left(e^{-\beta H}\right)$. 

\subsubsection{Dynamical quantities}\label{sec:NRG_dynamical_quantities}
Dynamical susceptibilities are computed for impurity and endpoint-electron operators from the retarded response $\chi''_{AB}(\omega,T) = \mathrm{Im}\,\chi^R_{AB}(\omega,T)$ where 
$
\chi^R_{AB}(t)
    =
    i\Theta(t)\langle [A(t),B(0)]\rangle_T\,.
$
Equivalently, the dissipative response is evaluated from the finite-temperature Lehmann representation using all states
\begin{equation}
    \chi''_{AB}(\omega,T)
    =
    \pi
    \sum_{m,n}
    (p_n-p_m)
    \langle n|A|m\rangle
    \langle m|B|n\rangle
    \delta\!\left(\omega-(E_m-E_n)\right),
\end{equation}
where $p_n = \frac{e^{-\beta E_n}}{Z}$. The static susceptibility used to locate the critical response is obtained from the Kramers--Kronig relation
\begin{equation}
    \chi'_O(0)
    =
    \frac{2}{\pi}
    \int_0^\infty d\omega\,
    \frac{\chi''_{OO}(\omega,T)}{\omega}.
\end{equation}

The finite-temperature QFI of a generator $O$ is reconstructed from the dissipative response according to eq. \ref{eq:qfi_response}.
The normalization of the NRG spectra is calibrated by the exact zero-temperature identity
$
    F_Q[|\psi_0\rangle,O]
    =
    4\,\mathrm{Var}_{0}(O),
$
where $\mathrm{Var}_{0}(O)$ denotes the variance in the ground state $|\psi_0\rangle$. For collective generators $ O_\eta = O_1+\eta O_2$, $\eta=\pm$, the corresponding dissipative response is assembled from the local response matrix as
\begin{equation}
    \chi''_{O_\eta O_\eta}
    =
    \chi''_{11}
    +
    \chi''_{22}
    +
    \eta
    \left(
        \chi''_{12}
        +
        \chi''_{21}
    \right),
\end{equation}
where $\chi''_{ij}$ denotes the response between $O_i$ and $O_j$. This decomposition is used for the impurity-spin generators $S^z_1+\eta S^z_2$, the endpoint-electron spin generators $s^z_{e,1}+\eta s^z_{e,2}$, and the endpoint charge generators $n_{e,1}+\eta n_{e,2}$. The spin-resolved endpoint density is decomposed as
$
    n^\uparrow_{e,i}
    =
    \frac{1}{2}n_{e,i}
    +
    s^z_{e,i},
$
so that
\begin{equation}
    n^\uparrow_{e,1}+\eta n^\uparrow_{e,2}
    =
    \frac{1}{2}
    \left(n_{e,1}+\eta n_{e,2}\right)
    +
    \left(s^z_{e,1}+\eta s^z_{e,2}\right).
\end{equation}
In the SU(2)-symmetric and non-magnetized state, the charge--spin cross response vanishes because the charge operator is a spin scalar while $s^z$ is the $z$ component of a spin vector. Therefore the spin-resolved density response is obtained from
\begin{equation}
    \chi''_{n^\uparrow_\eta n^\uparrow_\eta}
    =
    \frac{1}{4}
    \chi''_{n_\eta n_\eta}
    +
    \chi''_{s^z_\eta s^z_\eta},
\end{equation}
and the corresponding QFI satisfies
\begin{equation}
    F_Q[\rho_T,n^\uparrow_\eta]
    =
    \frac{1}{4}
    F_Q[\rho_T,n_\eta]
    +
    F_Q[\rho_T,s^z_\eta],
\end{equation}
with $n_\eta=n_{e,1}+\eta n_{e,2}$ and $s^z_\eta=s^z_{e,1}+\eta s^z_{e,2}$.

\begin{figure*}[htb!]
 \centering
 \begin{adjustbox}{center}
 %1
   \includegraphics[width=1\columnwidth]{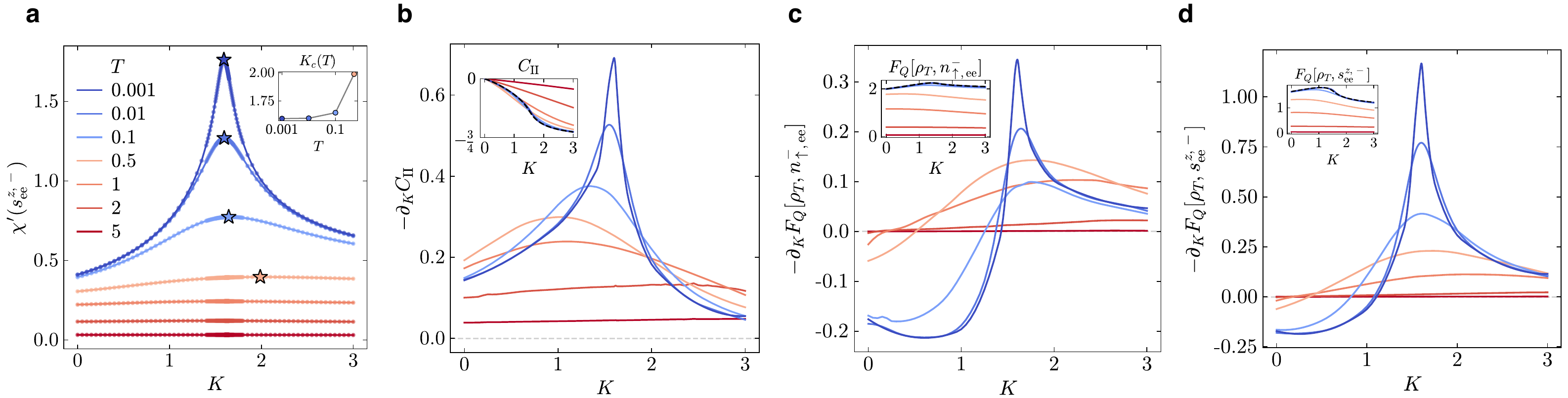} %2
 \end{adjustbox}
\caption{\textbf{Finite-temperature NRG results across the impurity critical region.}
\textbf{a}, Static susceptibility $\chi^{\prime}(s_{\mathrm{ee}}^{z,-})$ as a function of the inter-impurity exchange $K$ at fixed $J=2$ and temperatures $T=0.001$--$5$.
Stars mark the susceptibility maxima, and the inset shows the corresponding temperature-dependent peak $K_{c}(T)$.
\textbf{b}, Negative derivative of the impurity-impurity correlator, $-\partial_K C_{\mathrm{II}}$; the inset shows $C_{\mathrm{II}}$. The black lines denote $T=0$ DMRG data.
\textbf{c,d}, Negative derivatives of the QFI associated with the spin-resolved electron-density imbalance, $-\partial_K F_Q[\rho_T,n_{\uparrow,\mathrm{ee}}^{-}]$ (\textbf{c}), and the local electron-spin imbalance, $-\partial_K F_Q[\rho_T,s_{\mathrm{ee}}^{z,-}]$ (\textbf{d}); the corresponding undifferentiated QFI is shown in the insets.
The low-temperature responses develop pronounced features near the critical coupling, which broaden and are progressively suppressed with increasing temperature.}
\label{fig:NRG_additional_data}
\end{figure*}

Figure~\ref{fig:NRG_additional_data}(a) shows that the antisymmetric electron-spin susceptibility develops a pronounced low-temperature maximum near the critical coupling $K_c$ identified from the critical response to $S^{z,-}_{\rm II}$. With increasing temperature, the maximum broadens, decreases and shifts, defining the crossover scale shown in the inset. The same crossover appears in $-\partial_K C_{\mathrm{II}}$, see Fig.~\ref{fig:NRG_additional_data}(b), while the low-temperature NRG correlator approaches the zero-temperature DMRG result.
The derivatives of the electron-density and electron-spin QFIs in Fig.~\ref{fig:NRG_additional_data}c and d also develop characteristic structures near the critical region. The electron-spin QFI displays the sharper peak and follows the susceptibility most closely, identifying the antisymmetric local spin channel as the more sensitive probe. All features are progressively suppressed with increasing temperature, but remain visible over a finite temperature range as signatures of the quantum-critical crossover.

\subsubsection{Convergence of calculations}
\begin{figure*}[htb!]
 \centering
 \begin{adjustbox}{center}
   \includegraphics[width=1\columnwidth]{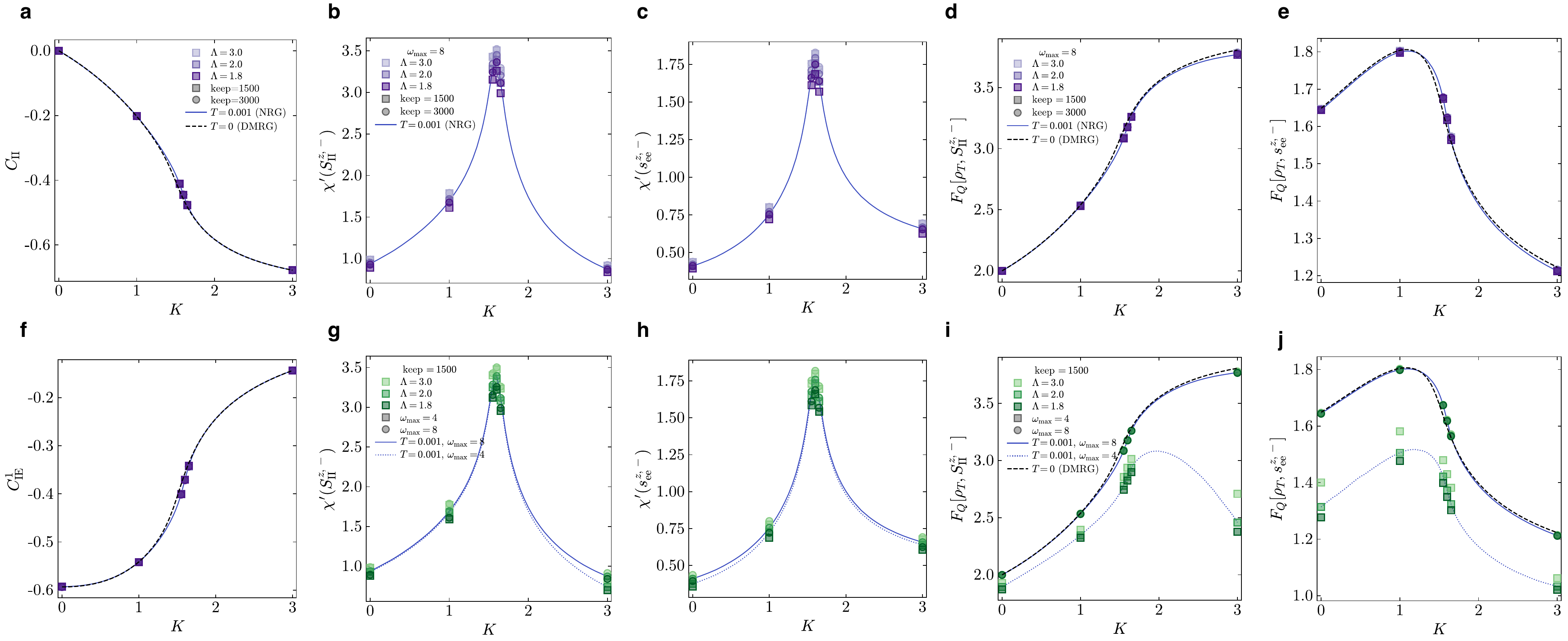} %2
 \end{adjustbox}
\caption{\textbf{Convergence of the finite-temperature NRG calculations.}
Observables as a function of the inter-impurity exchange $K$ at fixed $J=2$ and $T=0.001$.
\textbf{a,f}, Impurity--impurity correlator $C_{\mathrm{II}}$ and correlator $C_{\mathrm{IE}}^{1}$ between the impurity and the adjacent conduction-electron site, respectively.
\textbf{b,g}, Static susceptibility $\chi^{\prime}(S_{\mathrm{II}}^{z,-})$ of the staggered impurity-spin operator.
\textbf{c,h}, Static susceptibility $\chi^{\prime}(s_{\mathrm{ee}}^{z,-})$ of the staggered spin operator on the conduction-electron sites adjacent to the impurities.
\textbf{d,i}, Thermal quantum Fisher information $F_Q[\rho_T,S_{\mathrm{II}}^{z,-}]$ of the staggered impurity-spin generator.
\textbf{e,j}, Thermal quantum Fisher information $F_Q[\rho_T,s_{\mathrm{ee}}^{z,-}]$ of the staggered adjacent-electron spin generator.
In \textbf{a--f}, the marker shading distinguishes logarithmic discretizations $\Lambda=1.8$, $2.0$ and $3.0$, whereas the marker shape distinguishes truncation cutoffs $N_{\mathrm{keep}}=1500$ and $3000$; the dynamical observables in \textbf{b--e} are evaluated with the spectral cutoff $\omega_{\max}=8$.
In \textbf{g--j}, $N_{\mathrm{keep}}=1500$ is fixed, the same values of $\Lambda$ are compared, and the marker shape distinguishes $\omega_{\max}=4$ and $8$.
Solid blue curves denote the production NRG calculations with $\Lambda=2$, $N_{\mathrm{keep}}=1500$ and $\omega_{\max}=8$, whereas dotted blue curves in \textbf{g--j} show calculations with $\omega_{\max}=4$.
Black dashed curves in \textbf{a,d--f,i,j} show the corresponding zero-temperature DMRG results.
The data collapse under variations of $\Lambda$ and $N_{\mathrm{keep}}$ demonstrates convergence with respect to logarithmic discretization and many-body truncation, while the comparison in \textbf{g--j} shows that the smaller spectral cutoff $\omega_{\max}=4$ is insufficient and that $\omega_{\max}=8$ reproduces the zero-temperature DMRG reference.}
\label{fig:NRG_convergence}
\end{figure*}

The NRG calculations presented in the main text use a logarithmic discretization parameter $\Lambda=2$, $N_z=8$ interleaved discretization meshes for $z$ averaging, a truncation threshold of $N_{\rm keep}=1500$ retained many-body multiplets, and a rescaled energy cutoff $E_{\rm keep}=10$. The conduction band has half-width $D=2t=2$, given by the local density of states at the end of the tight-binding chain, so that $J$ and $K$ are measured in units of $t$. The Wilson chains are iterated down to the lowest energy scale $T_{\min}=\min(T/100,\,10^{-3})$, i.e.\ $T_{\min}=10^{-5}$ for the $T=0.001$ data shown here. Dynamical response functions are obtained within the full-density-matrix (FDM) NRG approach and evaluated on a logarithmic frequency mesh extending from $\omega_{\min}=10^{-7}$ to $\omega_{\max}=8$, with a ratio $r_\omega=1.01$ between consecutive frequency points, giving $N_\omega\simeq3.7\times10^{3}$ mesh points for both signs of $\omega$ combined; note that $\omega_{\max}$ is four times the band half-width, so the mesh extends well beyond the band edge. The discrete NRG spectra are accumulated in $N_{\rm bins}=1000$ histogram bins and broadened using a log-Gaussian kernel of width $\alpha=0.3$, supplemented at low frequencies by a Gaussian kernel of width $\gamma=0.2$. Here, $\Lambda$ controls the logarithmic discretization of the conduction-electron baths, $N_z$ is the number of shifted discretization meshes, $N_{\rm keep}$ is the maximum number of retained many-body multiplets, $E_{\rm keep}$ is the corresponding rescaled-energy cutoff, and $T_{\min}$ sets the lowest Wilson-chain scale reached in the calculation. Convergence is tested by varying $\Lambda$, repeating selected calculations with the larger truncation threshold $N_{\rm keep}=3000$, and reducing the upper frequency cutoff from $\omega_{\max}=8$ to $4$, while keeping all remaining parameters fixed at their production values.  

Figure~\ref{fig:NRG_convergence} demonstrates that the observables used in the main text are stable under the discretization, truncation and spectral-cutoff checks shown here. The static correlators $C_{\mathrm{II}}$ and $C_{\mathrm{IE}}^{1}$ in panels a and f are unchanged on the scale of the figure when the logarithmic discretization is varied over $\Lambda=1.8$--$3.0$ or when representative calculations are repeated with $N_{\rm keep}=3000$ instead of the production value $N_{\rm keep}=1500$; the corresponding relative deviations are below $0.3\%$. The low-temperature NRG results also closely follow the independent $T=0$ DMRG curves. At $\omega_{\max}=8$, the impurity- and electron-spin QFIs in panels d and e exhibit the same collapse under variations of $\Lambda$ and $N_{\rm keep}$, again to better than $0.3\%$, and agree closely with the corresponding DMRG results. The largest residual dependence occurs in the heights of the narrow susceptibility maxima in panels b and c, which increase by $2.8\%$ and $3.1\%$ respectively on going from $N_{\rm keep}=1500$ to $3000$; the overall $K$ dependence is nevertheless unchanged. We note that the convergence checks are performed on a coarse subset of six $K$ values and therefore constrain the magnitude of the susceptibilities rather than the precise location of their maxima.

Panels g-j isolate the dependence on the upper spectral cutoff at fixed $N_{\rm keep}=1500$. Reducing the cutoff to $\omega_{\max}=4$ preserves the critical peak in the static susceptibilities, but produces visible quantitative deviations and substantially underestimates both QFIs. In contrast, the results obtained with $\omega_{\max}=8$ are nearly independent of $\Lambda$ and reproduce the zero-temperature DMRG values throughout the investigated range of $K$, showing that the spectral weight relevant to the QFI integral is captured. 

The derivatives with respect to $K$ are obtained by analytically differentiating a smoothing cubic spline fitted to each temperature series to avoid the amplification of numerical noise on the dense, non-uniform $K$ grid. The spline is the smoothest curve $\tilde{f}$ satisfying $\sum_i [\,f(K_i)-\tilde{f}(K_i)\,]^2 \le s$ with smoothing strength $s = \lambda\, n\, \sigma^2$, where $n$ is the number of grid points, $\lambda = 10$, and $\sigma = 1.4826\,\mathrm{MAD}(\Delta^2 f)/\sqrt{6}$ is a Gaussian noise estimate from the median absolute deviation of the second differences. For the present data this criterion is satisfied with $\approx 10$ spline knots across $K \in [0,3]$, so the fit is substantially stiffer than the $153$-point grid it is applied to; increasing $\lambda$ from $10$ to $100$ shifts the extracted peak positions by less than $0.01$ in $K$ and their heights by less than $4\%$. Since markers on a differentiated curve would denote spline evaluations rather than computed data points, the derivative panels show the fitted curve alone, whereas the undifferentiated panels also mark the individual $K$ points.

\subsection{Four-spin model}
The four-spin model provides a minimal strong-coupling description of the local degrees of freedom in the two-impurity Kondo problem by retaining only the two impurity spins and the two conduction-electron spins closest to them. The Hamiltonian is given by
\begin{equation}
    H_{\rm 4-spin}
    =
    J\left(\mathbf S_1\cdot \mathbf s_1+\mathbf S_2\cdot \mathbf s_2\right)
    +
    K\,\mathbf S_1\cdot \mathbf S_2 \,.
\end{equation}
For consistency, we use $C_{\rm II}$ for the impurity-impurity correlator and $C_{\rm IE}$ for the correlator between electron-spin and impurity-spin, and $S(A_{\rm imp})$ and $S(A_L)$ for the entropy of the central impurity spins with respect to the conduction electron spins and the central cut entanglement, respectively. Similarly, we keep the same notation for the QFI of $S^{z,-}_{\rm II}$.

\subsubsection{Zero-temperature ground state}
\begin{figure*}[htb!]
 \centering
 \begin{adjustbox}{center}
   \includegraphics[width=0.75\columnwidth]{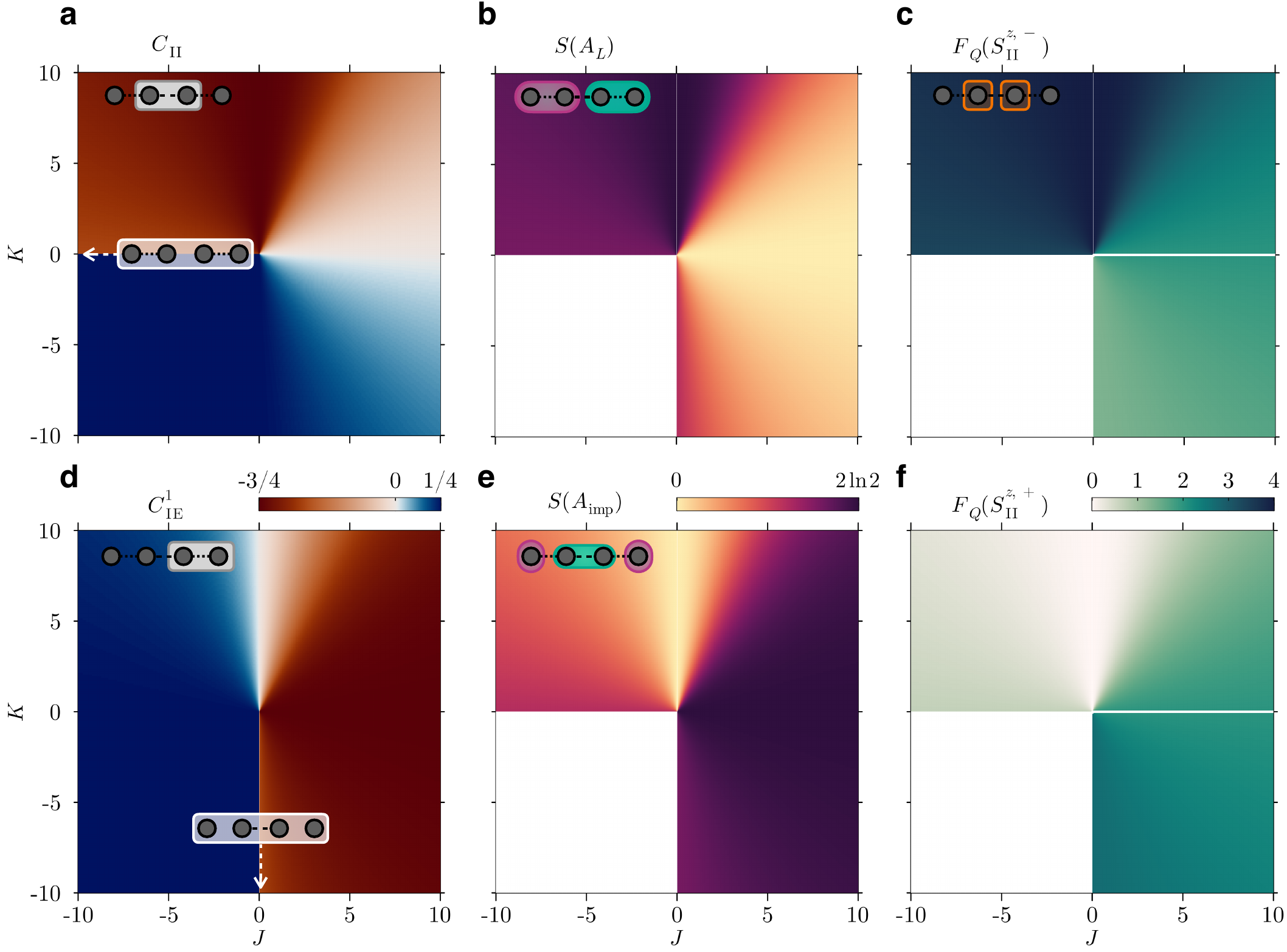}
 \end{adjustbox}
    \caption{\textbf{Ground-state phase diagrams of the four-spin model.}
    \textbf{a}, Impurity-impurity correlator $C_{\rm II}$.
    \textbf{b}, Entanglement entropy $S(A_L)$ of the local impurity--electron dimer. White quadrant corresponds to degenerate ground state sector. 
    \textbf{c}, Quantum Fisher information $F_Q(S^{z,-}_{\mathrm{II}})$ for the staggered impurity-spin generator detects entanglement for $K>0$. White line denotes contour at $F_Q = 2$.
    \textbf{d}, Local impurity-electron correlator $C_{\rm IE}$.
    \textbf{e}, Entropy of the impurity subsystem $S(A_{\mathrm{imp}})$.
    \textbf{f}, Quantum Fisher information $F_Q(S^{z,+}_{\mathrm{II}})$ for the uniform impurity-spin generator.}
\label{fig:SI_four_spin_model_phase_diagrams}
\end{figure*}

Fig. \ref{fig:SI_four_spin_model_phase_diagrams} contains the zero-temperature ground state phase diagrams of the four-spin model. For large antiferromagnetic $J>0$, the ground state is locally close to two electron--impurity singlets, $C_{\rm II} \to 0$ and $C_{\rm IE} \to -3/4$, as shown in panels (a) and (d), respectively. This state competes with antiferromagnetic $K>0$, which favors an impurity singlet, $C_{\rm II} \to -3/4$. Thus, in the sector $J,K>0$, the system has a crossover between a two-Kondo-singlet regime for $J\gg K$ and an impurity-singlet regime for $K\gg J$. In the isolated four-spin model this is not a true phase transition: the two singlet configurations have the same total spin and hybridize, producing an avoided level crossing instead of a non-analytic transition in the full model. In the sector $J>0$ and $K<0$, the impurities are driven into a triplet with $C_{\rm II}\to 1/4$, while the antiferromagnetic $J$ still tends to screen this composite impurity spin through negative electron-impurity correlations, $C_{\rm IE}<0$. For $J<0,K>0$, the impurity coupling favors an impurity singlet state with $C_{\rm II} = -3/4$, but the ferromagnetic electron-impurity exchange favors local triplets with $C_{\rm IE}>0$. Finally, for $J<0$ and $K<0$, all couplings are ferromagnetic, and the four spins lock into a maximal-spin object with $S_{\rm tot}=2$ and $C_{\rm IE}, C_{\rm II}\to 1/4$.

\begin{figure*}[htb!]
 \centering
 \begin{adjustbox}{center}
   \includegraphics[width=1\columnwidth]{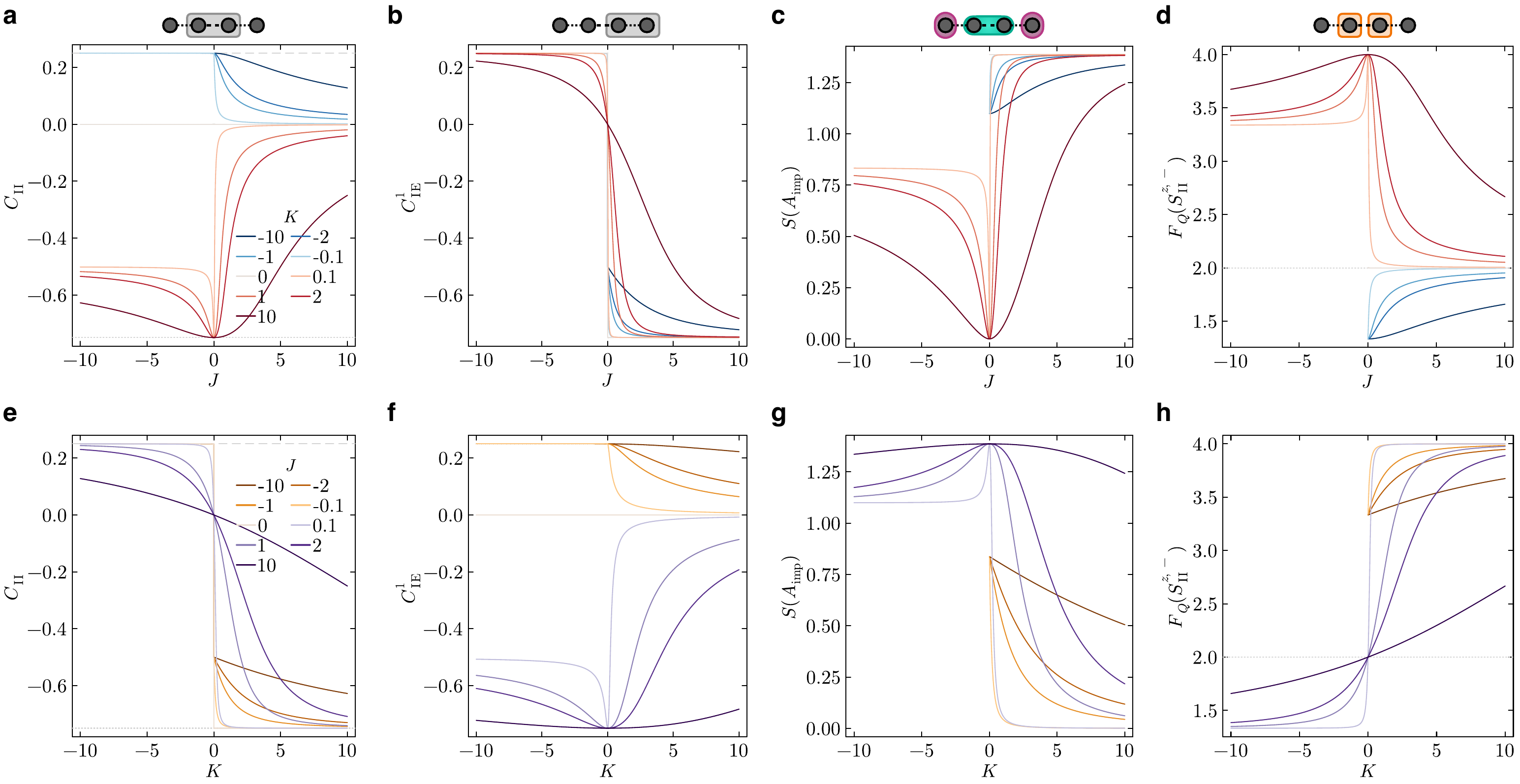}
 \end{adjustbox}
\caption{\textbf{Ground-state observables of the four-spin model.}
Correlators, entanglement entropies and QFI as a function of $J$ at fixed $K$ (top row) and of $K$ at fixed $J$ (bottom row), corresponding to horizontal and vertical cuts across the phase diagrams in Fig.~\ref{fig:SI_four_spin_model_phase_diagrams}, respectively.
\textbf{a,e}, Impurity--impurity correlator $C_{\mathrm{II}}$, distinguishing impurity-singlet formation for antiferromagnetic $K>0$ from triplet formation for ferromagnetic $K<0$.
\textbf{b,f}, Local impurity--electron correlator $C_{\mathrm{IE}}$, which resolves antiferromagnetic singlet formation for $J>0$ and ferromagnetic polarization for $J<0$.
\textbf{c,g}, Entanglement entropy $S(A_{\mathrm{imp}})$ of the impurity subsystem. 
\textbf{d,h}, QFI $F_Q(S^{z,-}_{\mathrm{II}})$ for the staggered impurity-spin generators. The dotted line denotes the bipartite-entanglement threshold $F_Q=2$.}
\label{fig:SI_four_spin_model_linecuts}
\end{figure*}

For $K=0$ and $J>0$, the Hamiltonian factorizes into two independent impurity-electron dimers, and the ground state is a product of two local singlets, $C_{\rm II} = 0$, see Fig. \ref{fig:SI_four_spin_model_linecuts}a. The left-right entanglement entropy between the two local dimers is zero, $S(A_L) = 0$, whereas the entanglement between the impurities and the rest of the system $S(A_{\rm imp})$ is maximal, see Fig. \ref{fig:SI_four_spin_model_linecuts}(c). In contrast, for $J=0$ and $K>0$, the impurities form an isolated antiferromagnetic Heisenberg dimer with $C_{\rm II} = -3/4$, as shown in Fig. \ref{fig:SI_four_spin_model_linecuts}f, whereas for $J=0$ and $K<0$, they form a triplet with $C_{\rm II} = 1/4$. The point $J=K=0$ is therefore not a unique physical limit, since all four spins are free and the ground state is highly degenerate. For $J,K>0$, the exact singlet-sector ground-state energy is
\begin{equation}
    E_0(J,K)
    =
    -\frac{K}{4}
    -\frac{J}{2}
    -\frac{1}{2}\sqrt{K^2-2KJ+4J^2},
\end{equation}
and the Hellmann--Feynman theorem gives
\begin{equation}
    C_{\rm II}
    =
    \frac{\partial E_0}{\partial K}
    =
    -\frac{1}{4}
    +
    \frac{J-K}{2\sqrt{K^2-2KJ+4J^2}},
\end{equation}
which reproduces all limiting cases.
Thus, $K>0$ lifts the degeneracy of the free impurities and selects the impurity-singlet direction at $J=0$, explaining the abrupt difference between the $K=0$ and $K>0$ curves. For fixed $K>0$, the strong-$J$ limit is
\begin{equation}
    C_{\rm II}
    \simeq
    -\frac{3K}{16J}
    \to 0
    \qquad (J\gg K),
\end{equation}
corresponding to two independent local impurity-electron singlets. In the four-spin model, the QFI of the difference and sum of impurity spins, as shown in Fig. \ref{fig:SI_four_spin_model_phase_diagrams} (c) and (f), respectively, allows entanglement to be certified in all sectors. The antisymmetric combination follows the central cut entanglement $S(A_L)$, compare panel (b) to (c). $F_Q(S^{z,-}_{\rm II})$ certifies left–right entanglement between the two impurity–bath subsystem for all values of $J$ if $K>0$, as shown in the linecuts in Fig. \ref{fig:SI_four_spin_model_linecuts}, which can be related to the entropy $S(A_{\rm imp})$. The symmetric combination, shown in Fig. \ref{fig:SI_four_spin_model_phase_diagrams}(f), certifies entanglement in the sector $J>0,K<0$.

In the presence of a degenerate ground-state manifold spanned by orthonormal states $\{|\psi_a\rangle\}_{a=1}^g$, expectation values of observables are not unique if one selects a single pure state, since different linear combinations within the degenerate subspace can give different results. For ordinary correlators, however, one can define a basis-independent ground-space averaged expectation value
\begin{equation}
    \langle \mathcal{O}\rangle_{\rm av}
    =
    \frac{1}{g}\sum_{a=1}^g
    \langle \psi_a|\mathcal{O}|\psi_a\rangle
    =
    \frac{1}{g}\Tr(P_0 \mathcal{O}),
\end{equation}
where $P_0=\sum_{a=1}^g|\psi_a\rangle\langle\psi_a|$ is the projector onto the degenerate ground-state subspace. This corresponds to evaluating $\mathcal{O}$ in the equal-weight mixed ground-state density matrix $\rho_0=P_0/g$. Thus, for spin-spin correlators, averaging over the degenerate ground states gives a well-defined ground-space averaged correlator and avoids arbitrary dependence on the eigenbasis chosen by the numerical diagonalization. In contrast, pure-state entanglement entropies and the pure-state QFI formula are not uniquely defined in a degenerate ground space. Therefore, the phase diagrams in \ref{fig:SI_four_spin_model_phase_diagrams}(b,c) and (e,f) are empty in the sector $K,J<0$.

\subsubsection{Finite temperature}
In units of $k_{\mathrm B}=1$, the probability of an eigenstate $\ket{n}$ with energy $E_n$ at temperature $T$ is $p_n(T)= e^{-E_n/T}/Z$, where the thermal partition function is $Z=\sum_n e^{-E_n/T}$.
For numerical stability at low temperatures, we equivalently evaluate the weights as $e^{-(E_n-E_0)/T}$, where $E_0$ is the ground-state energy. The thermal expectation value of an observable $\mathcal{O}$ is then obtained as
\begin{equation}
    \langle \mathcal{O}\rangle_T
    =
    \operatorname{Tr}\!\left(\rho_T\mathcal{O}\right)
    =
    \sum_n p_n(T)\bra{n}\mathcal{O}\ket{n},
\end{equation}
where $\rho_T = e^{-H_{\mathrm{4-spin}}/T}/Z$ is the thermal density matrix. The sixteen eigenstates decompose into two singlets, three triplet multiplets, and one quintet multiplet. Their energies are
\begin{equation}
\begin{split}
E_{{\rm S},\pm}
= &
-\frac{J}{2}
-\frac{K}{4}
\pm
\frac{1}{2}
\sqrt{4J^2-2JK+K^2}, \\
E_{{\rm T},0}
= &
-\frac{J}{2}
+\frac{K}{4}, \\
E_{{\rm T},\pm}
= &
-\frac{K}{4}
\pm
\frac{1}{2}
\sqrt{J^2+K^2}, \\
E_{\rm Q}
= &
\frac{J}{2}
+\frac{K}{4} .
\end{split}
\end{equation}
The two singlet levels are non-degenerate, each triplet level has degeneracy three, and the quintet has degeneracy five. The four-spin partition function is therefore
\begin{equation}
Z_{\rm 4}
={}
e^{-E_{{\rm S},-}/T}
+
e^{-E_{{\rm S},+}/T}
+
3e^{-E_{{\rm T},0}/T}
\
+
3e^{-E_{{\rm T},-}/T}
+
3e^{-E_{{\rm T},+}/T}
+
5e^{-E_{\rm Q}/T}.
\end{equation}
Introducing the free energy $F_{\rm 4}=-T\ln Z_{\rm 4}$, the finite-temperature correlators follow directly from the Hellmann-Feynman theorem:
\begin{equation}
    C_{\rm II}(T)
=
-T\partial_K\ln Z_{\rm 4}
\end{equation}
and
\begin{equation}
    C_{\rm IE}(T)
=
-\frac{T}{2}
\partial_J\ln Z_{\rm 4} .
\end{equation}

\begin{figure*}[htb!]
 \centering
 \begin{adjustbox}{center}
   \includegraphics[width=1\columnwidth]{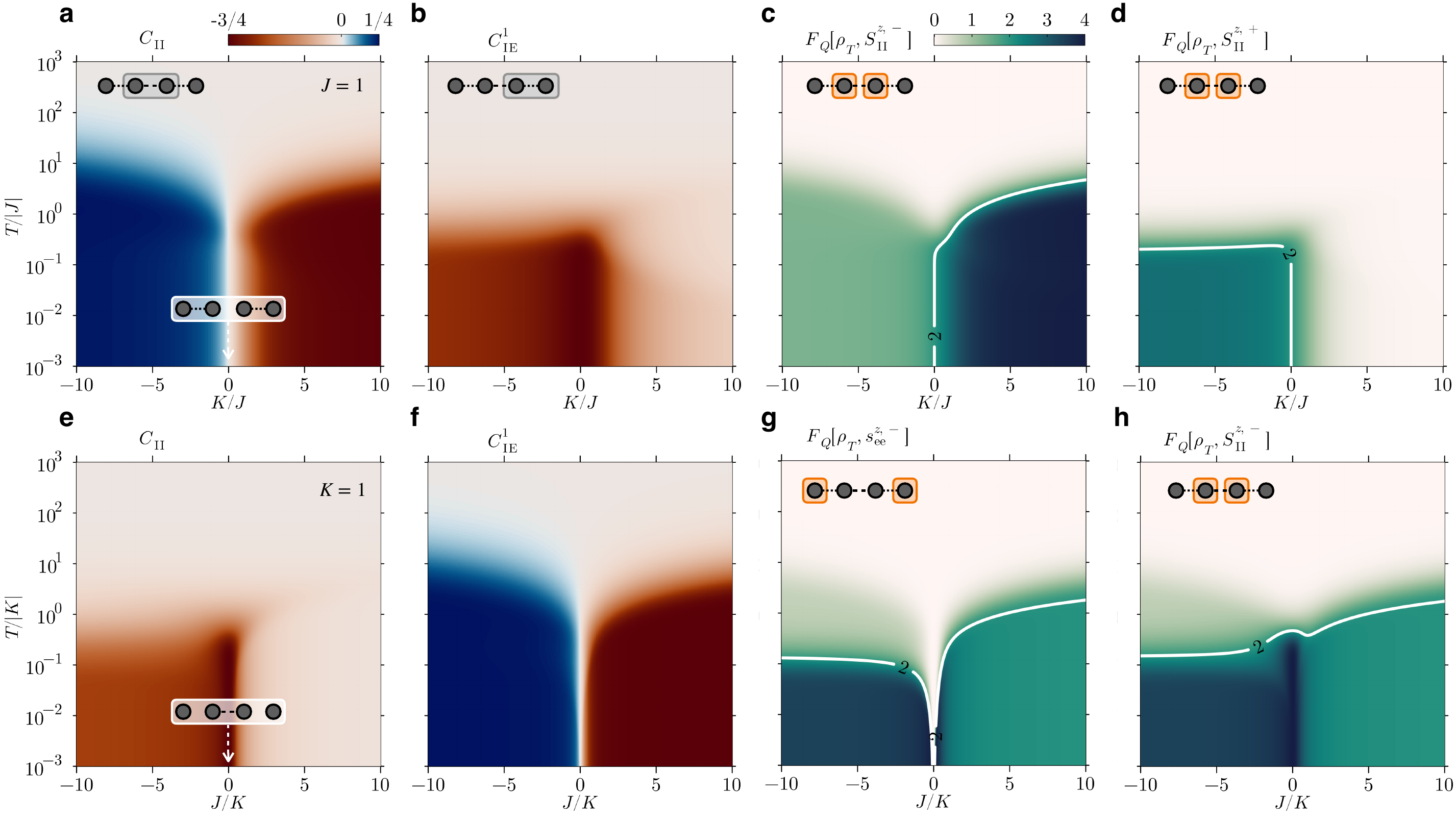}
 \end{adjustbox}
\caption{\textbf{Thermal phase diagrams of the four-spin model.}
Correlators and QFI as a function of temperature and coupling ratio. The top row shows $K/J$ at fixed $J=1$, while the bottom row shows $J/K$ at fixed $K=1$.
\textbf{a,e}, $C_{\mathrm{II}}(T)$ indicates predominantly singlet (triplet) states for $K>0$ ($K<0$) up to $T/J \sim 10$. At $K=0$, the impurities are decoupled from one another. For $K=1$ (\textbf{e}), the impurities are in a singlet state with $C_{\rm II} < 0$ for $J>0$ up to $T/K \sim 1$ and decouple at larger temperatures. 
\textbf{b,f}, $C_{\mathrm{IE}}(T)$ shows that impurities are AFM screened for $K<0$ up to $T/J \sim 1$. This region is stable up to $K/J \sim 1$. For $K=1$ (\textbf{f}), the impurity-electron coupling becomes zero for $J\rightarrow 0$. The impurity-electron dimers form a singlet (triplet) for $J>0$ ($J<0$).
\textbf{c,d}, $F_Q(S^{z,-}_{\mathrm{II}})$ and $F_Q(S^{z,+}_{\mathrm{II}})$ certify entanglement for $K>0$ and $K<0$, respectively. White contours denote $F_Q=2$. Increasing temperature suppresses correlations and reduces the regions in which the QFI witnesses bipartite entanglement. Note that the antisymmetric generator is stable up to higher temperatures compared to the symmetric one. 
\textbf{g}, $F_Q(S^{z,-}_{\rm EE})$ becomes zero for $J\rightarrow 0$. \textbf{h}, $F_Q(S^{z,-}_{\mathrm{II}})$ witnesses entanglement at up to $T/K \sim 1$ even as $J/K \to 0$, while $F_Q(S^{z,-}_{\rm ee})$ does not.}
\label{fig:SI_four_spin_model_thermal_phase_diagrams}
\end{figure*}

\begin{figure*}[htb!]
 \centering
 \begin{adjustbox}{center}
   \includegraphics[width=1\columnwidth]{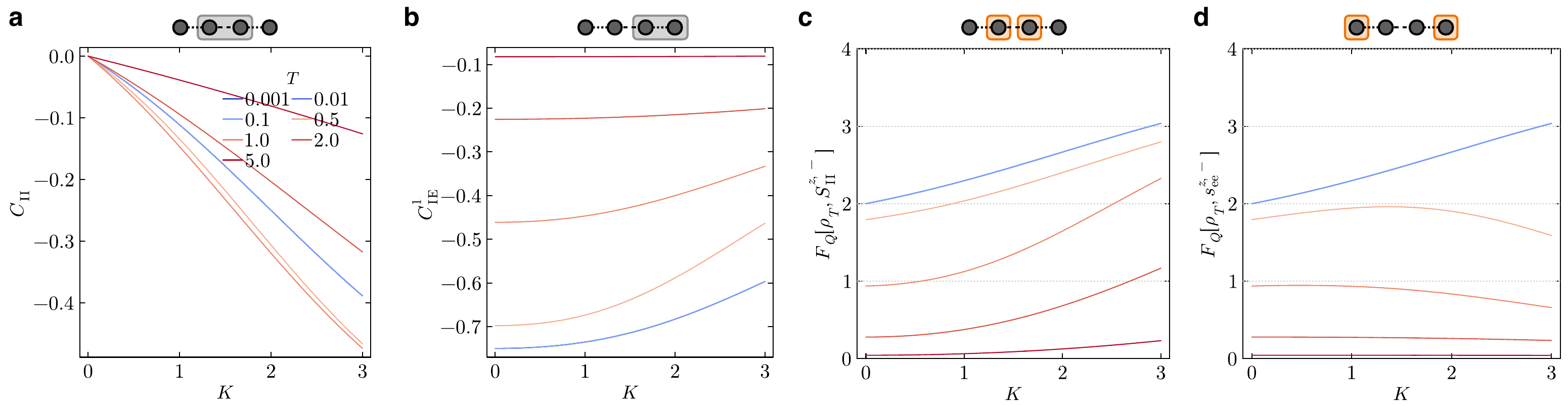}
 \end{adjustbox}
\caption{\textbf{Finite-temperature correlations and quantum Fisher information in the four-spin model.}
Thermal-state results as a function of the inter-impurity exchange $K$ at $J=2$ and temperatures $T=0.001$--$5$.
\textbf{a}, Impurity--impurity correlator $C_{\mathrm{II}}$.
\textbf{b}, Local impurity--electron correlator $C_{\mathrm{IE}}^{1}$.
\textbf{c}, QFI of the staggered impurity-spin operator, $F_Q[\rho_T,S_{\mathrm{II}}^{z,-}]$.
\textbf{d}, QFI of the staggered electron-spin operator, $F_Q[\rho_T,s_{\mathrm{ee}}^{z,-}]$.
Increasing temperature progressively weakens the spin correlations and suppresses the QFI, illustrating the thermal degradation of the corresponding spin fluctuations.}
\label{fig:SI_four_spin_model_thermal_linecuts}
\end{figure*}

For $K=0$, the Hamiltonian factorizes into two independent impurity-electron Heisenberg dimers. The local electron-impurity singlets are thermally destroyed on the scale $T\sim J$ within the truncated model. Similarly, for $J=0$, we have an isolated dimer, which is stable up to $T \sim K$.
For generic $J$ and $K$, the isolated four-spin model remains finite-dimensional and therefore has no finite-temperature phase transition. For $J,K>0$, the zero-temperature avoided level crossing between the two impurity-electron singlet and impurity-singlet configurations is further broadened by thermal occupation of the excited multiplets. Defining the smallest excitation gap as
\begin{equation}
\Delta_{\rm 4}(J,K)
=
\min_{\nu\neq0}
\left[
E_{\nu}(J,K)-E_0(J,K)
\right],
\end{equation}
thermal corrections to low-temperature observables are exponentially suppressed for $T\ll\Delta_{\rm 4}$ and become appreciable once $T\gtrsim\Delta_{\rm 4}$. 
In the opposite high-temperature limit, all sixteen states are populated approximately equally, leading to
\begin{equation}
    C_{\rm II}(T)
    =
    -\frac{3K}{16T}
    +
    \mathcal O(T^{-2})
\end{equation}
and 
\begin{equation}
    C_{\rm IE}(T)
    =
    -\frac{3J}{16T}
    +
    \mathcal O(T^{-2}).
\end{equation}

At finite temperature, the density matrix is mixed, and the QFI becomes
\begin{equation}
F_Q[\rho_T,\mathcal O]
=
2
\sum_{m,n}
\frac{(p_m-p_n)^2}{p_m+p_n}
\left|
\bra{m}
\mathcal O
\ket{n}
\right|^2
\leq
4\operatorname{Var}{\rho_T}(\mathcal O).
\end{equation}
The inequality is generally strict because the variance contains classical thermal fluctuations in addition to quantum fluctuations.

For $J=2$ and $K>0$, the ground state is the lower singlet $E_{{\rm S},-}$, and the leading thermal corrections arise predominantly from occupation of the lowest triplet multiplet $E_{{\rm T},-}$. At higher temperatures, the remaining triplets, the upper singlet and ultimately the quintet also contribute. This evolution is illustrated in Fig. \ref{fig:SI_four_spin_model_thermal_linecuts}. Increasing $K$ strengthens the antiferromagnetic impurity correlation $C_{\rm II}$ while reducing the magnitude of the local impurity--electron correlation $C_{\rm IE}^{1}$, see Fig. \ref{fig:SI_four_spin_model_thermal_linecuts}(a,b), reflecting the redistribution of correlations from the two local impurity--electron bonds towards the impurity dimer. The corresponding impurity- and electron-spin QFIs increase with $K$ at low temperature, but are progressively suppressed and smoothed by thermal population of the excited multiplets, see Fig. \ref{fig:SI_four_spin_model_thermal_linecuts}(c,d). These trends are qualitatively consistent with the NRG results, which likewise show thermal suppression and broadening of the spin response. The finite four-spin model, however, produces only a smooth local crossover and cannot reproduce the sharp critical enhancement generated by the extended conduction baths in NRG.

%%

%TC:endignore

\end{document}